\documentclass[fleqn,usenatbib]{mnras}

\usepackage{newtxtext,newtxmath}

\usepackage[T1]{fontenc}

\DeclareRobustCommand{\VAN}[3]{#2}
\let\VANthebibliography\thebibliography
\def\thebibliography{\DeclareRobustCommand{\VAN}[3]{##3}\VANthebibliography}

\usepackage{graphicx}	
\usepackage{amsmath}	
\usepackage{caption}
\usepackage{multicol}
\usepackage{multirow}
\usepackage{array}
\usepackage{bm}

\title[On the assembly state of DM haloes]{On the choice of the most suitable indicator for the assembly state of dark matter haloes through cosmic time}

\author[Vallés-Pérez et al.]{
David Vallés-Pérez,$^{1}$\thanks{E-mail: david.valles-perez@uv.es}
Susana Planelles,$^{1,2}$
Óscar Monllor-Berbegal,$^{1}$
Vicent Quilis$^{1,2}$
\\
$^{1}$Departament d’Astronomia i Astrofísica, Universitat de València, E-46100 Burjassot (València), Spain\\
$^{2}$Observatori Astronòmic, Universitat de València, E-46980 Paterna (València), Spain
}

\date{Accepted XXX. Received YYY; in original form ZZZ}

\pubyear{2023}

\begin{document}
\label{firstpage}
\pagerange{\pageref{firstpage}--\pageref{lastpage}}
\maketitle

\begin{abstract}
The dynamical state and morphological features of galaxies and galaxy clusters, and their high-redshift precursors, are tightly connected with their assembly history, encoding crucial information about the formation and evolution of such cosmic structures. 
As a first step towards finding an optimal indicator of the assembly state of observed structures, we use a cosmological simulation of a moderate volume to critically examine the best definition of an indicator that is able to discriminate dark matter haloes undergoing mergers and/or strong accretion from haloes experimenting a relaxed evolution. 
Using a combination of centre offset, virial ratio, mean radial velocity, sparsity and ellipticity of the dark matter halo, we study how the thresholds on these parameters, as well as their relative weights, should evolve with redshift to provide the best classification possible. 
This allows us to split a sample of haloes in a totally relaxed, a marginally relaxed and an unrelaxed subsamples. 
The resulting classification strongly correlates with the merging activity obtained from the analysis of complete merger trees extracted from whole simulation data.
The results on how the different indicators depend on redshift and halo mass, and their optimal combination to better match the true assembly history of haloes, could constitute relevant hints to find a suitable set of indicators applicable to observational data. 
\end{abstract}




\section{Introduction}
\label{s:intro}

Deeply interwoven through a complex network of filaments and sheets, dark matter (DM) haloes are bound, diffuse structures which result from the gravitational collapse of the primordial density fluctuations and a hierarchical merging history \citep{Zeldovich_1970, Press_1974, Gott_1975}. DM haloes constitute the fundamental building blocks of the large-scale structure (LSS) of the Universe, and host their baryonic counterparts that we observe over the electromagnetic spectrum (see, for instance, \citealp{Planelles_2015}, for a review). At the galactic scale, the current theories of galaxy formation typically assume DM haloes to be virialised \citep[e.g.,][]{White_1978}, although this does not necessarily hold for each galactic DM halo; while, at larger masses (at the galaxy cluster scale), most DM haloes are still expected to be in the process of virialisation, since they are the latest objects to have assembled \citep[e.g.,][for a review on galaxy cluster formation]{Kravtsov_2012}.

However, the dynamical state of individual haloes is tightly connected to their assembly history and, in particular, to the presence of mergers and the accretion rates in the last one or few dynamical times. A merger or a period of intense accretion usually triggers many morphological and dynamical disturbances in the halo (asphericity, higher velocity dispersions, abundance of substructures, changes to the internal structure, etc.), which gradually fade away once the assembly episode is over (see, for example, \citealp{Poole_2006} for a thorough analysis of the disturbances and the subsequent relaxation after a merger event at cluster scales).

Since dynamically relaxed and disturbed structures often present fundamentally different properties, a characterisation of the dynamical state of the sample of cosmic structures is often a necessary procedure in many analyses of very different natures, such as in studies about the geometry of the cosmic web \citep{Gouin_2021}, statistical properties of the population of galaxy clusters (scaling relations, mass functions, etc.; e.g., \citealp{Chen_2019, Seppi_2021}), hydrostatic mass bias \citep{Nelson_2014, Biffi_2016, Angelinelli_2020}, turbulence \citep{Vazza_2017, Valdarnini_2019, Valles_2021cpc, Valles_2021, Simonte_2022}, or galactic environments \citep{Kuchner_2022}, just to mention a few. 

Even though we usually define haloes using the virial radius prescription of \cite{Eke_1996} and \cite{BryanNorman_98}, based on the spherical collapse model, this does not imply that, in general, three-dimensional haloes (non necessarily spherical, in a full-cosmological, i.e., not isolated environment) defined this way are necessarily in virial equilibrium. While in simulations one can access the whole temporal evolution of the objects, and thus recover the assembly history of the halo under study in order to assess the dynamical state, this is not possible in observations. Thus, for the sake of a more direct comparison with observational works, simple schemes for characterising the dynamical state using halo properties at a given time are usually involved in many analyses.

For the time being, most works have relied on placing a threshold on some halo property expected to correlate with the dynamical state, in order to split the relaxed and unrelaxed subsamples. Perhaps, the most direct of such indicators is the \textit{virial ratio}, usually defined as $\eta \equiv 2T/|W|$, where $T$ is the intrinsic kinetic energy of the halo and $W$ is its gravitational potential energy. $\eta$ would be expected to be 1 for an isolated system in a steady state. However, different works have found different thresholds to best suite their particular classification \citep[e.g.,][see also the discussion in \citealp{Cui_2017}]{Shaw_2006, Neto_2007, Knebe_2008}. Similarly, there is debate about the necessity of including a surface tension term to account for the fact that haloes are not isolated \citep{Poole_2006, Shaw_2006, Knebe_2011}. Another frequently used indicator, both in simulations and observations, is the centre offset, which quantifies the departure from smoothness and spherical symmetry of the matter distribution, and serves as an indicator of substructure \citep{Crone_1996}. In practice, however, there are many possibilities regarding the choice of centres \citep[see][]{Cui_2016} and how to set the thresholds (cf. \citealp{dOnghia_2007, Maccio_2007}). Additionally, in observations the centre offset may depend crucially on the orientation, posing an additional challenge. Last, other authors have used the fraction of mass in substructures as a measure of the dynamical unrelaxedness of a DM halo (e.g., \citealp{Neto_2007}; cf. other recently suggested approaches, e.g. \citealp{Kimmig_2022}). While the election of this magnitude is well-motivated, the mass contained in substructures in simulated haloes depends critically on numerical resolution and the precise definition of the substructure extent (see, e.g, the discussion in \citealp{Valles_2022}), making this criterion less comparable.

Since it is difficult that a single property can reflect the complex picture of the dynamical state of a halo, many recent studies have used combinations of these indicators, either by considering as relaxed the haloes which simultaneously fulfil several relaxation criteria \citep{Neto_2007, Biffi_2016}, or by defining some combined indicator \citep{Haggar_2020, ZhangB_2021, DeLuca_2021}. Finally, other metrics of the dynamical state are based on the X-ray morphology, such as the centroid shift $\omega$ \citep{Mohr_1993}, or the power ratio, $P_3/P_0$ \citep[][see also the review of \citealp{Rasia_2013} on X-ray morphological estimators for galaxy clusters]{Buote_1995}; or more sophisticated ones such as those involving Fourier analyses of the fluctuations in mass and X-ray maps \citep{Cerini_2022}, or the expansion of the Compton $y$-maps in Zernike polynomials \citep{Capalbo_2021}.

However, in most of the previous works, the parameters being used and, especially, the thresholds imposed on them have been tuned in a somewhat empirical way. This has lead to variations in the criteria from work to work, even though the underlying idea is kept. 
Furthermore, a possible redshift evolution of these thresholds or of their very relevance has been devoted marginal attention, either because the studies were focused on a particular cosmic epoch or because it had been implicitly assumed that these criteria should not evolve with redshift. 

In this work, we intend to critically examine a set of possible indicators of the assembly state, all of which can be obtained from the complete three-dimensional information in simulations, and develop a criterion which accommodates redshift-dependent thresholds and the possibility that different indicators have more or less relevance at different cosmic epochs. We note the reader that, while in the following we may refer to the \textit{dynamical} state of haloes, our main focus is oriented towards the dynamical disturbances associated to the assembly history of haloes (i.e., the presence of merger events or episodes of strong accretion; rather than a more general sense of dynamical unrelaxedness which could include, e.g., the presence of substructures even when they are not associated to a merger episode, since they have an impact on properties such as the hydrostatic equilibrium).

The rest of the manuscript is organised as follows. In Sec. \ref{s:methods}, we introduce our simulation, halo sample and the methodology that we employ for setting the thresholds and relative weights of the different dynamical state indicators. Our resulting criterion is presented in Sec. \ref{s:results}, including the analysis of the mass dependence of our results and a validation of our method with a different simulation. Finally, we discuss the applicability of our results in Sec. \ref{s:conclusions}. Appendix \ref{s:appA} contains the fitting formulae for the thresholds and weights applicable for massive haloes.

\section{Methods}
\label{s:methods}
The results reported in this paper have been extracted from the analysis of a $\Lambda$CDM cosmological simulation tracking the coupled evolution of baryons and DM. We describe the relevant details of the simulation in Sec. \ref{s:methods.simu}, then cover the halo catalogues and merger tree elaboration in Sec. \ref{s:methods.haloes}, and discuss how do we compute the dynamical state indicators in Sec. \ref{s:methods.indicators}. Finally, we introduce our classification strategy in Sec. \ref{s:methods.classification}.

\subsection{The simulation}
\label{s:methods.simu}
The haloes we analyse in this paper are extracted from a numerical simulation run with \texttt{MASCLET} \citep{Quilis_2004, Quilis_2020}, a (magneto-)hydrodynamics and $N$-Body code primarily designed for cosmological applications. For evolving the DM component, which is the primary focus of this work, \texttt{MASCLET} implements a multilevel Particle-Mesh (PM) scheme \citep{Hockney_Eastwood_1988}, which takes advantage of the adaptive-mesh refinement (AMR) strategy \citep{Berger_Colella_1989} to gain spatial, temporal and force resolution.

We have simulated a periodic, cubic ($L=100 \, h^{-1} \, \mathrm{Mpc}$) domain, under the assumption of a flat, $\Lambda$CDM cosmology specified by the matter density parameter $\Omega_m=0.31$ ($\Omega_\Lambda = 1 - \Omega_m$), baryon density parameter $\Omega_b = 0.048$, and Hubble parameter $h \equiv H_0 / (100 \, \mathrm{km \, s^{-1}}) = 0.678$. The initial conditions stem from a realisation of the primordial gaussian random field assuming a spectral index $n_s = 0.96$ and an amplitude yielding $\sigma_8 = 0.82$, and are set up at redshift $z_\mathrm{ini}=100$ using a CDM transfer function \citep{Eisenstein_1998}. The values selected for the cosmological parameters are consistent with the latest results reported by \cite{Planck_2020}.

A first simulation is run at low resolution, using a fix grid of $N_x^3 = 256^3$ cells and the same number of equal-mass particles. This is used to identify the Lagrangian regions in the initial conditions which will evolve into dense structures by $z=0$, and mapping them with enhanced numerical resolution already at $z_\mathrm{ini}$. We use this approach to establish three nested levels of initial conditions, resulting in a best mass resolution of $1.48 \times 10^7 \, M_\odot$.

Using these high-resolution initial conditions, the simulation is evolved again using AMR based on gas/DM overdensities, converging flows, and Jeans length criteria, achieving a peak resolution of $\Delta x_8 = 2.3 \, \mathrm{kpc}$ at the maximum ($\ell = n_\ell \equiv 8$) level of refinement. While the baryonic component is not the primary focus of this work, the simulation includes gas cooling, but no other baryonic effect or feedback mechanism.

\subsection{Halo catalogue and merging history}
\label{s:methods.haloes}
For each snapshot of the simulation, we have identified the DM haloes using the public halo finder \texttt{ASOHF} \citep{Planelles_2010, Knebe_2011, Valles_2022}\footnote{\url{https://github.com/dvallesp/ASOHF}.}, which is based on the spherical-overdensity definition and uses the virial radius (according to the prescription of \citealp{BryanNorman_98}) to delimit the extent of the haloes that are not substructure.

After determining the halo catalogues, these are linked in between snapshots using the merger tree code presented by \citet[their section 2.6.2]{Valles_2022}, which identifies all the haloes at a given code output which have contributed to an object in a following one, allowing to skip an arbitrary number of snapshots, if necessary. Using it, we determine the main evolutionary line of each halo, as well as the presence and characterisation of mergers. 

Following \cite{Planelles_2009}, \cite{Chen_2019}, and \cite{Valles_2020}, we have classified each merger event in the sample as either a major merger (if the mass ratio, $M_\mathrm{min}/M_\mathrm{max}$, between the two haloes involved exceeds $1/3$), or a minor merger ($1/3 > M_\mathrm{min}/M_\mathrm{max} \geq 1/10$). Mergers below a mass ratio of $1/10$ are disregarded. The merger time is determined as the moment in which the centre of the infalling (the least massive) halo crosses the virial boundary of the host (the most massive) halo.

\subsubsection{Fiducial classification: assembly history of the haloes}
\label{s:methods.haloes.fiducial}

In order to determine the optimal thresholds on the dynamical state indicators (see below, Sec. \ref{s:methods.indicators} and therein), we compare with a reference, or \textit{fiducial}, classification of the dynamical state based on the full assembly history of haloes (i.e., the presence of past or ongoing mergers, as well as the accretion rates). 

As a tentative classification of the unrelaxedness induced by a merger event, we will assume that a typical halo remains in a disturbed state for one dynamical time after a major merger, or half a dynamical time after a minor merger, with the dynamical time $\tau_\mathrm{dyn}$ being defined as

\begin{equation}
    \tau_\mathrm{dyn}(z) \equiv \frac{1}{\sqrt{G \rho}} = \frac{1}{\sqrt{G \rho_B(z) \Delta_\mathrm{vir}(z)}},
    \label{eq:dyntime}
\end{equation}

\noindent with $G$ the gravitational constant, $\rho$ the density of the halo, $\rho_B(z)$ the background matter density and $\Delta_\mathrm{vir}(z)$ the virial overdensity \citep{BryanNorman_98}.

While the choice of the timespan is a crude approximation, it responds to the fact that many works have shown that the disturbance triggered by a minor merger is, in general terms, smaller than the effect of a major merger, both for the dark and for the baryonic components \citep{Planelles_2009, Yu_2014, Valles_2020, Zhang_2021}. In practical terms, since $\tau_\mathrm{dyn}(z)$ varies strongly with redshift and reaches considerable fractions of the age of the Universe, especially at low redshift, we choose to define the number of dynamical times between two moments, $t_1$ and $t_2$, as in \cite{Jiang_2016} and \cite{Wang_2020}:

\begin{equation}
   N_\tau(t_1, t_2) = \int_{t_1}^{t_2} \frac{\mathrm{d} t}{\tau_\mathrm{dyn}(z)}
   \label{eq:dyntime_integral}
\end{equation}

Additionally, it might be the case that a halo is accreting strongly, but without undergoing any significant merger (either physically or due to the finite resolution of a simulation). Thus, we also consider as unrelaxed, for the purpose of the fiducial classification, any halo which has assembled more than $50\%$ of their mass in the last dynamical time.

\begin{figure}
    \centering
    \includegraphics[height=0.33\textheight]{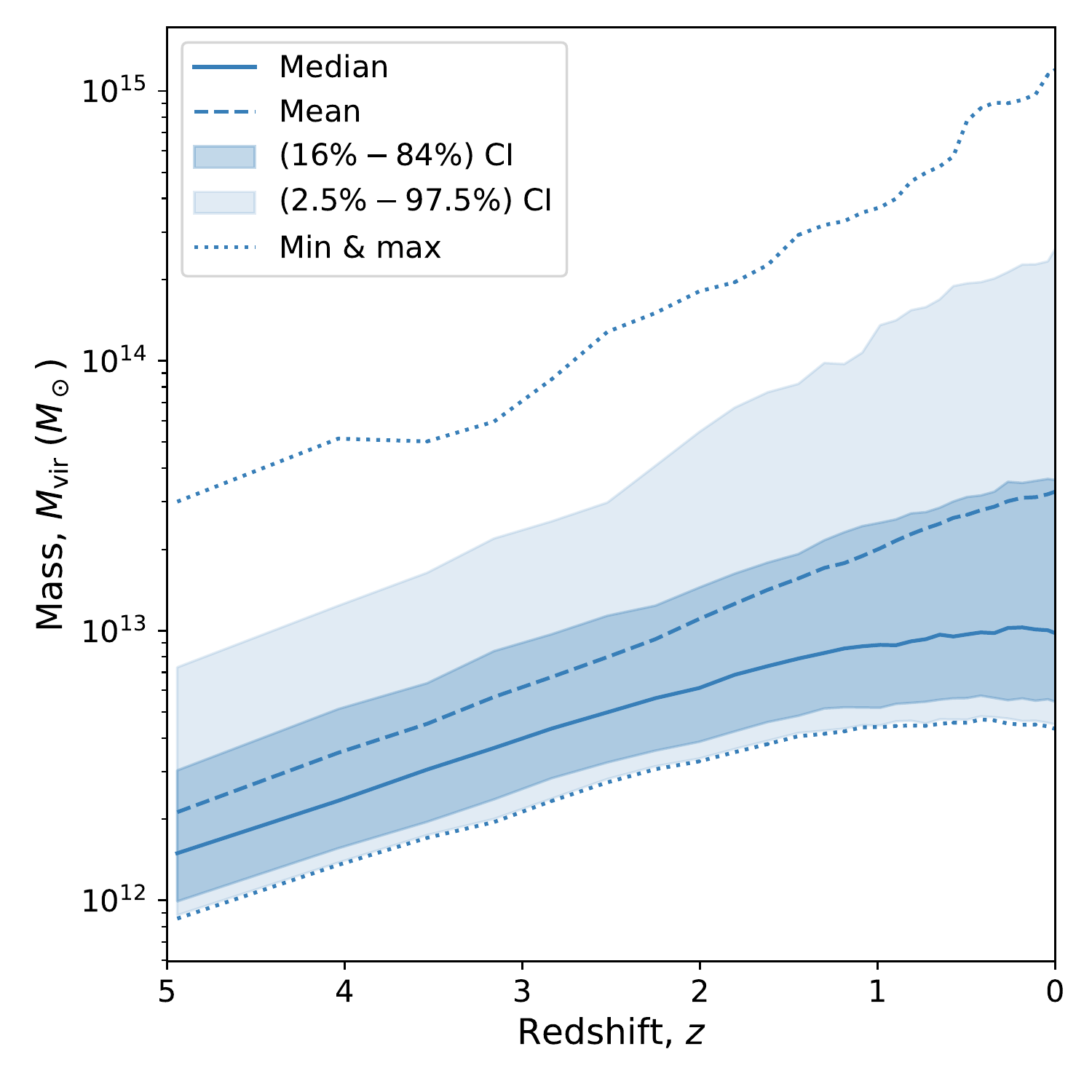}
    \caption{Evolution of the distribution of halo masses in our sample. The solid line indicates the median mass of the sample, with the dark and light shaded regions enclosing the $16\%-84\%$ (dark blue) and $2.5\%-97.5\%$ (light blue) confidence intervals (CIs) around it, respectively. The dashed line corresponds to the mean mass, while the dotted lines correspond to the maximum and minimum masses.}
    \label{fig1}
\end{figure}

For the analyses within this work, all the 28 snapshots of the simulation since redshift $z=5$ are considered. We select the 1000 most massive haloes at each epoch, and discard all those which cannot be reliably traced back in time for at least one $\tau_\mathrm{dyn}(z)$. We show, in Fig. \ref{fig1}, the redshift evolution of the median mass in the sample (solid line), together with shaded regions enclosing the confidence intervals corresponding to the $16\%-84\%$ (dark blue) and $2.5\%-97.5\%$ (light blue) percentiles of the distribution of masses. The dotted lines mark the maximum mass (upper line) and the minimum mass, or mass limit (lower line) in the sample at each time. Thus, the mass limit in our sample evolves from $\sim 10^{12} M_\odot$ at $z=5$ to $\sim 4.5 \times 10^{12} M_\odot$ at $z=0$. The wide redshift interval considered in this study includes from the cluster-, group- and massive galaxy-sized haloes at $z\simeq 0$, to the DM counterpart of galaxies and the progenitors of low-redshift clusters at the high-redshift end.

\begin{figure}
    \centering
    \includegraphics[height=0.33\textheight]{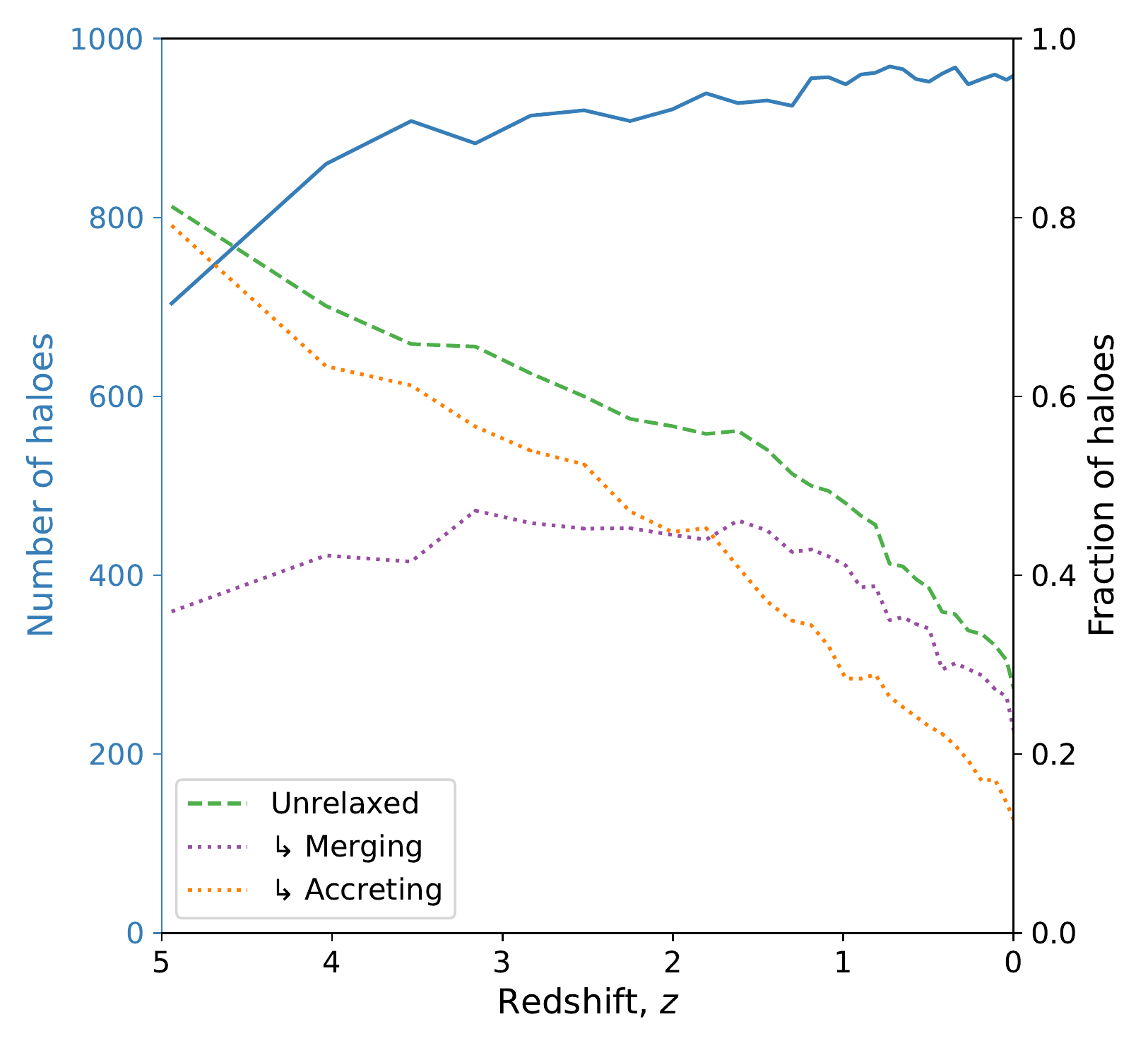}
    \caption{Fiducial classification of dynamical states of the halo sample. The blue line represents the number of haloes selected per snapshot, according to the left axis. The dashed/dotted lines, according to the right axis, describe the evolution with redshift of the fraction of unrelaxed haloes (green), which can have been labelled as such due to mergers (purple) or strong accretion/mass growth (orange).}
    \label{fig2}
\end{figure}

The results of the fiducial classification are summarised in Fig. \ref{fig2}, where we show the number of haloes which are finally considered at each snapshot (blue line, referring to the axis on the left). Only at high redshift ($z \gtrsim 3$), a large fraction ($10\%$ to $25\%$) of the preliminary haloes get discarded because they cannot be tracked back in time for at least one dynamical time. The fraction of unrelaxed haloes according to the fiducial classification (green, dashed line; referring to the axis on the right) varies from $\sim 80\%$ to $\sim 30\%$ through the considered redshift interval. Purple and orange, dotted lines show the number of haloes, as a fraction of the total, which are unrelaxed due to either the condition on recent mergers or the condition on the accretion rate, respectively. Most of the low-redshift haloes which are labelled unrelaxed are undergoing mergers, while at high redshift the cause for unrelaxedness is more usually a high level of smooth accretion. This may be due to several reasons, amongst which we can mention the higher density in the vincinity of haloes at high redshift, or resolution limitations of the simulation (i.e., at high redshift, a halo may be accreting small, underresolved structures, which are therefore not accounted as mergers).

\subsection{Indicators for the assembly state}
\label{s:methods.indicators}

Many possible proxies for the dynamical and assembly state of a DM halo, or their corresponding baryonic structure (e.g., a galaxy or a galaxy cluster) have been proposed in the literature \citep[see, for instance,][just to cite a few]{Cole_1996, Crone_1996, Shaw_2006, Haggar_2020, ZhangB_2021}. While simulations allow to access the complete three-dimensional picture, the lack of the whole information in observations (due to, e.g., projection or the inability to observe the dark component, or even the plasma out to large radii) requires that, generally, different quantities are used for assessing the dynamical state in simulations and in observations \citep{Rasia_2013, DeLuca_2021, Yuan_2022}. While comparison with observations is crucial and will be dealt with in future work, here we shall focus on the dynamical state indicators extracted from the complete, three-dimensional data in simulations as a first step. Unless otherwise specified, all quantities below are referred to the virial volume.

\paragraph*{Centre offset.} The centre offset is usually defined as the distance between two different choices of centre, in units of some aperture radius (typically, the virial radius of the halo, $R_\mathrm{vir}$). Many examples for the choices of centre pair exist in the literature, such as centre of mass (CM) vs. density peak \citep{Baldi_2017} or CM vs. potential minimum \citep{Biffi_2016}, extracted from the three-dimensional description in simulations, or the morphological offset of the BCG location vs. X-ray surface brightness peak \citep{Rosetti_2016}, amongst many others. We address the interested reader to \cite{Cui_2016}, who compare many different choices of observable for defining the centre of galaxy clusters.

In this work, we have tested the three possible combinations between the minimum of gravitational potential (defined as the location of the most-bound DM particle, as obtained by \texttt{ASOHF} and described in detail in \citealp{Valles_2022}), the DM density peak, and the DM centre-of-mass. We find that the most robust results are obtained for the Peak-CM pair. Therefore, we defined the centre offset parametre as:

\begin{equation}
    \Delta_r = \frac{\left|\bm{r}_\mathrm{peak,DM} - \bm{r}_\mathrm{CM,DM}\right|}{R_\mathrm{vir}}
    \label{eq:centreoffset}
\end{equation}

\paragraph*{Virial ratio.} For a gravitational system in steady state, the virial theorem predicts $2T + W - E_s = 0$, where $T$ is the kinetic energy, $W$ is the gravitational binding energy, and $E_s$ is the surface energy term \citep{Chandrasekhar_1961}. Neglecting the surface term, the virial ratio is usually defined as

\begin{equation}
    \eta \equiv \frac{2T}{|W|},
    \label{eq:virial_ratio}
\end{equation}

\noindent and it is expected that $\eta \to 1$ for isolated systems. However, haloes are not generally isolated systems, and therefore there is not a good \textit{a priori} reason to drop the surface term in the virial theorem. Thus, many works define the virial ratio as $\eta' = (2T-E_s)/|W|$ \citep{Shaw_2006}, while others claim that the surface term overcorrects the virial ratio \citep{Power_2012}. As the latter, we find that correcting the virial ratio by the surface term wipes out the correlation with merging activity, and thus we shall use the definition in Eq. \ref{eq:virial_ratio} in the remainder of this work.

\paragraph*{Mean radial velocity.} In a relaxed object, we do not expect important changes in the radial structure, while an unrelaxed system will experience significant disturbances as it settles down to equilibrium. This motivates the consideration of the mean radial velocity of DM particles,

\begin{equation}
    \langle v_r \rangle_\mathrm{DM} = \frac{\sum_i m_i v_{r,i}}{\sum_i m_i},
\end{equation}

\noindent being $m_i$ the mass of the $i$-th DM particle, and $v_{r,i}$ its radial velocity relative to the halo reference frame. In practical terms, we scale this quantity by the circular velocity at the virial radius, $V_\mathrm{circ,vir} \equiv \sqrt{{G M_\mathrm{vir}}/{R_\mathrm{vir}}}$, and define the corresponding normalised indicator as

\begin{equation}
    \langle \widetilde{v_r} \rangle_\mathrm{DM} \equiv \frac{\left|\langle v_r \rangle_\mathrm{DM}\right|}{V_\mathrm{circ,vir}}.
\end{equation}

\paragraph*{Sparsity.} Systems which have experienced recent significant mergers tend to display shallower central density profiles due to the disturbance caused by the infalling halo, and thus are less concentrated. Many works \citep[e.g.,][]{Neto_2007, Wang_2020} have pointed out the relation between the time spanned since the last major merger and halo concentration, $c_\mathrm{vir} = R_\mathrm{vir} / R_\mathrm{s}$, being $R_\mathrm{s}$ the scale radius of the \cite{Navarro_1997} profile (or the radius where the logarithmic slope of the DM density profile equals $-2$). 

More recently, sparsity has been suggested as a non-parametric alternative to concentration, which reduces the scatter with halo mass \citep{Balmes_2014, Corasaniti_2018}, and has also been found to correlate with the timing since the last relevant merger \citep{Richardson_2022}. While sparsity is generally defined as the quotient between the masses at different spherical overdensities, we find that the one maximising the correlation with merging activity is

\begin{equation}
    s_{200c,500c} \equiv \frac{M_{200c}}{M_{500c}}.
    \label{eq:sparsity}
\end{equation}

\paragraph*{Ellipticity.} DM haloes are generally triaxial \citep{Frenk_1988, Knebe_2006}, with significant scatter in halo shape at a given mass and redshift. Many recent studies have pointed out at the correlation between triaxiality and/or ellipticity of the halo shape and the formation history of a halo, with relaxed haloes tending to be rounder \citep{Chen_2019, Lau_2021}. 

We define the overall shape of the DM halo by finding the eigenvalues of the shape tensor, defined as

\begin{equation}
    S_{\alpha\beta} = \sum_i m_i \frac{r_{i,\alpha} r_{i,\beta}}{r_i^2},
    \label{eq:shape_tensor}
\end{equation}

\noindent which are proportional to the semiaxes squared. The positions, $\bm r_i$, are relative to the cluster centre (defined as the location of the density peak), and we choose to normalise them to be unit length to prevent the shape to be dominated by the particles in the outskirts of the halo. Note this corresponds to the E2 method introduced by \cite{Zemp_2011}. If $a$, $b$ and $c$ are the semiaxes sorted in non-increasing order, we define the ellipticity of the halo, $\epsilon$, as:

\begin{equation}
    \epsilon = 1 - \frac{c}{a}.
    \label{eq:ellipticity}
\end{equation}

\paragraph*{Other indicators not considered in this work.} Amongst the most widely used proxies for the dynamical state of DM haloes in the literature, we have not included the fraction of substructures, $f_\mathrm{sub}$, in this study (neither defined as the mass in substructures as a fraction of the host mass, nor as the ratio between the mass of the heaviest substructure and the host mass, as in \citealp{Cialone_2018}). While $f_\mathrm{sub}$ should naturally correlate with the assembly state (especially, with the merging state), its interpretation is very subtle due to several factors. First, there is not a unique way to define the extent of a subhalo, and differences amongst halo finders have a dramatic impact on the recovered masses of substructures (see \citealp[their figures 5 and 10]{Valles_2022}). In second place, the amount of substructure produced in simulations depends strongly, not only on resolution, but also on the numerical scheme employed to solve gravity. This introduces strong mass biases (while the most massive haloes in our simulation may host well-resolved substructure, haloes with less than a few ten thousands particles are likely to be substructure-deficient. These obscure dependencies with mass, resolution and numerical scheme limit our ability to consistently incorporate this indicator in our work. Simulations with enhanced resolution, capable of fully resolving rich substructure in our wide range of masses could be able to overcome this limitation of our work.

Regarding the indicators describing the shape of the mass distribution, while $\epsilon$ alone does not fully characterise the shape of an ellipsoid, we have not considered any additional parameter, such as triaxiality $T \equiv \frac{a^2-b^2}{a^2-c^2}$ \citep{Franx_1991}. While ellipticity measures directly the deviation from sphericity, which is expected during assembly episodes, the same is not true for triaxiality. As a matter of fact, triaxiality is undefined for spherical objects, and we do not find a clear reason to have a preference towards either prolateness/oblateness during mergers or strong accretion periods.

\subsection{Classification strategy}
\label{s:methods.classification}

\subsubsection{Redshift binning}
\label{s:methods.classification.redhisft_binning}

A total of 28 snapshots of the simulation, since $z=5$, are saved and used in this analysis. To augment the statistics, we have grouped the snapshots in several redshift bins, which are described in Table \ref{tab1}.\footnote{Not all bins contain the same number of snapshots (or haloes): higher redshift bins comprise less snapshots. While this may increase the scatter in our results at high redshift, grouping more snapshots together at high redshift would increase the systematic uncertainty due to stacking objects of more different epochs.}

\begin{table}
    \centering
    \caption{Summary of the redshift binning considered for the subsequent analyses. Each bin contains the haloes extracted from the $N_\mathrm{snaps}$ available with $z \in [z_\mathrm{min}, \, z_\mathrm{max}]$. The mean redshift of the $N_\mathrm{haloes}$ haloes in the bin is $\bar z$, with a fraction $f_\mathrm{unrelaxed}$ of them being unrelaxed (either merging or experiencing intense accretion) according to the fiducial classification. Note we report $\bar z$, instead of the median, because $z$ is not continuously distributed (at each redshift bin, there are only $N_\mathrm{snaps}$ different values of $z$).}
    \begin{tabular}{cc|cc|c|c}
         $z_\mathrm{min}$ & $z_\mathrm{max}$ & $N_\mathrm{snaps}$ & $\bar z$ & $N_\mathrm{haloes}$ & $f_\mathrm{unrelaxed}$  \\ \hline
         0 & 0.2 & 4 & 0.084 & 3828 & 0.309\\
         0.2 & 0.5 & 4 & 0.381 & 3830 & 0.360 \\
         0.5 & 0.75 & 3 & 0.651 & 2890 & 0.406 \\ 
         0.75 & 1.0 & 3 & 0.897 & 2871 & 0.468 \\
         1.0 & 1.5 & 4 & 1.253 & 3769 & 0.512 \\
         1.5 & 2.0 & 3 & 1.808 & 2788 & 0.562 \\
         2.0 & 3.0 & 3 & 2.536 & 2742 & 0.600 \\
         3.0 & 4.0 & 2 & 3.350 & 1791 & 0.657 \\ 
         4.0 & 5.0 & 2 & 4.443 & 1564 & 0.751
    \end{tabular}
    \label{tab1}
\end{table}

\subsubsection{Optimising the thresholds}
\label{s:methods.classification.thresholds}

In a first step, we place a threshold, $X_i^\mathrm{thr}$, for each of the dynamical state indicators, $X_i$, described in the previous section ($i=1,\dots,5$, for the five dynamical state indicators). This is performed independently at each redshift bin. To do so, we vary $X_i^\mathrm{thr}$ from the minimum to the maximum value of $X_i$ through the sample, and identify how well does $X_i^\mathrm{thr}$ separate the relaxed and the unrelaxed samples of the fiducial classification.

For each value of $X_i^\mathrm{thr}$, we compute two complementary metrics of the goodness of the classification\footnote{Note that the metrics introduced in Eqns. \ref{eq:efficiency_unrelaxed} and \ref{eq:efficiency_relaxed} also correspond, respectively, to the True Positive Rate (TPR) or sensitivity, and the True Negative Rate (TNR) or specificity in the usual jargon of binary classifications \citep[e.g.,][]{Fawcett_2006}. However, we choose this notation here for better readability.}, namely the efficiency in discriminating the unrelaxed haloes,

\begin{equation}
    \epsilon_\mathrm{unrelaxed} (X_i^\mathrm{thr}) = \frac{\text{\# of unrelaxed haloes properly identified}}{\text{\# of unrelaxed haloes (fiducial)}}
    \label{eq:efficiency_unrelaxed}
\end{equation}

\noindent and the efficiency in discriminating the relaxed haloes,

\begin{equation}
    \epsilon_\mathrm{relaxed} (X_i^\mathrm{thr}) = \frac{\text{\# of relaxed haloes properly identified}}{\text{\# of relaxed haloes (fiducial)}}.
    \label{eq:efficiency_relaxed}
\end{equation}

Out of all the possible values of $X_i^\mathrm{thr}$, we choose the one which maximises the product of both metrics, that is to say:

\begin{equation}
    \hat X_i^\mathrm{thr} = \mathrm{argmax}_{X_i^\mathrm{thr}} \left[  \epsilon_\mathrm{unrelaxed} (X_i^\mathrm{thr}) \cdot \epsilon_\mathrm{relaxed} (X_i^\mathrm{thr}) \right].
    \label{eq:choice_threshold}
\end{equation}

Since $\epsilon_\mathrm{unrelaxed}$ ($\epsilon_\mathrm{relaxed}$) can be thought, in a frequentist approach, as the probability of correctly identifying an unrelaxed (relaxed) halo as such, our choice of threshold in Eq. \ref{eq:choice_threshold} corresponds to picking the one which enhances the likelihood of correctly classifying both an unrelaxed and a relaxed halo, and thus serves as a compromise between too generous and too stringent thresholds.

\subsubsection{Totally relaxed, marginally relaxed and disturbed haloes}
\label{s:methods.classification.sample}

Once the final (redshift-dependent) thresholds, $\left\{ X_i^\mathrm{thr}(z) \right\}_{i=1}^5$, are established, any halo will be regarded as \textit{totally relaxed} if

\begin{equation}
    X_i < X_i^\mathrm{thr}(z) \qquad \forall i=1,\dots,5,
\end{equation}

\noindent that is, if it has a low value of all the dynamical state indicators (low centre offset, mean radial velocity and ellipticity, virial ratio and sparsity close to unity). This allows a very conservative definition of the most relaxed haloes.

However, it may be the case that a halo has a high value of one of the parametres, but is relaxed according to the rest. This might be the case for a variety of reasons, ranging from physical (e.g., a halo with high ellipticity due to a strong tidal field generated by the surrounding large-scale structure; \citealp{Chen_2016}) to numerical (e.g., underresolved haloes with higher sparsities, misidentification of the centre, etc.). Thus, we deal with all haloes not falling into the \textit{totally relaxed} category by defining a combined \textit{relaxedness} indicator, in the manner of \cite{Haggar_2020} (see also \citealp{Kuchner_2020, ZhangB_2021, Gouin_2022}) but adding weights which account for the fact that some dynamical state indicators can be more insightful than others at any given particular epoch.

\begin{equation}
\begin{aligned}
    \chi = 
    \left[
    w_1 \left(\frac{\Delta_r}{\Delta_r^\mathrm{thr}} \right)^2 + 
    w_2 \left(\frac{\eta-1}{\eta^\mathrm{thr}-1} \right)^2 + 
    w_3 \left(\frac{\langle \widetilde{v_r} \rangle_\mathrm{DM}}{\langle \widetilde{v_r} \rangle_\mathrm{DM}^\mathrm{thr}} \right)^2 + \right. \\
    \left. w_4 \left(\frac{s_{200c,500c}-1}{s_{200c,500c}^\mathrm{thr}-1} \right)^2 + 
    w_5 \left(\frac{\epsilon}{\epsilon^\mathrm{thr}} \right)^2
    \right]^{-1/2}
\end{aligned}
\label{eq:xi_dynstate}
\end{equation}

The weights, $\{w_i\}_{i=1}^5$, are normalised so that $\sum_{i=1}^5 w_i = 1$, and are fixed at each redshift bin to be proportional to the performance of their corresponding indicator in splitting the merging and non-merging subsamples of the fiducial classifications. In particular, we set $w_i \propto \epsilon_\mathrm{relaxed} \epsilon_\mathrm{unrelaxed} - 0.25$ (the absolute values being set by the closure relation $\sum_{i=1}^5 w_i = 1$). If, at a given redshift bin, $\epsilon_\mathrm{relaxed} \epsilon_\mathrm{unrelaxed} \leq 0.25$, we consider that the particular indicator is not meaningful and its weight is set to $w_i=0$.

A particular halo which does not belong to the \textit{totally relaxed category} will be classified as \textit{marginally relaxed} if $\chi \geq 1$, and \textit{disturbed} whenever $\chi < 1$. Additionally, this classification scheme can naturally handle missing data. For instance, if $s_{200c,500c}$ is missing (e.g., due to a low resolution not enabling to resolve $R_{500c}$), one can simply evaluate $\chi$ neglecting the sparsity term (and multiplying $\chi$ by a factor $\sqrt{1-w_4}$; or, alternatively, renormalising the weights after setting $w_4=0$).

\subsubsection{Redshift evolution of the thresholds and weights}
\label{s:methods.classification.redshift_evolution}

With the procedure outlined in Sec. \ref{s:methods.classification.thresholds} and \ref{s:methods.classification.sample}, we obtain a threshold and a weight for each dynamical state indicator at each of the redshift bins specified in Table \ref{tab1}. In order to obtain a continuous trend for each of these parameters, we fit them to polynomial functions of arbitrary degree.

First, we estimate the uncertainties in the thresholds ($X_i^\mathrm{thr}$) and weights ($w_i$) by computing the standard deviation of the distribution of these parametres obtained in 1000 bootstrap resamplings \citep{Efron_1979}. Then, we fit the redshift evolution of the given parameter to polynomial functions of increasing degree, until the $p$-value of the highest degree coefficient falls above $p=0.046$ (low significance), or the reduced chi-squared falls below 1 (indicating possible overfitting of the model). Fits are performed using least squares weighted to the inverse of the variance of each data point.

\section{Results}
\label{s:results}

\begin{figure}
    \centering
    \includegraphics[width=0.8\linewidth]{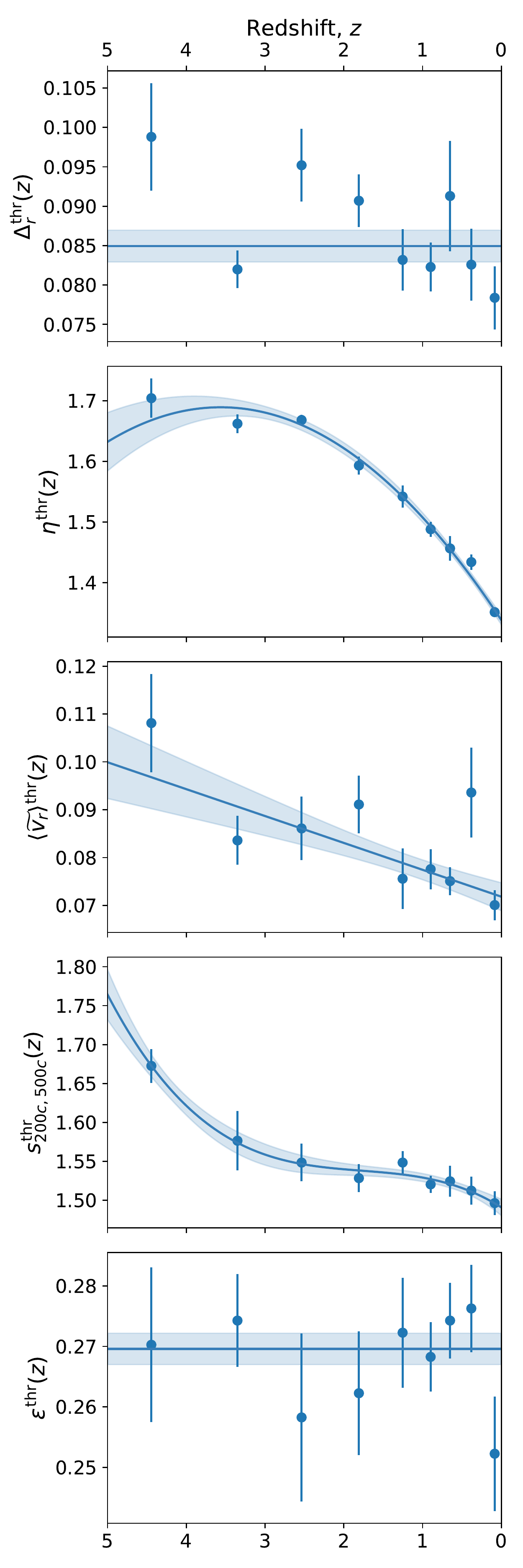}
    \caption{Redshift evolution of the thresholds on the dynamical state indicators. From top to bottom, the panels refer to the centre offset ($\Delta_r^\mathrm{thr}$), virial ratio ($\eta^\mathrm{thr}$), mean radial velocity ($\langle \widetilde{v_r} \rangle_\mathrm{DM}^\mathrm{thr}$), sparsity ($s_{200c,500c}^\mathrm{thr}$), and ellipticity ($\varepsilon^\mathrm{thr}$) thresholds. Dots correspond to the optimal threshold obtained within the redshift bin, with the error bars obtained by means of bootstrap resampling. Solid lines correspond to the best polynomial fits, with their (16-84)\% confidence interval as the shaded region.}
    \label{fig3}
\end{figure}

Following the procedure described in Sec. \ref{s:methods.classification.thresholds} and \ref{s:methods.classification.redshift_evolution} over the whole sample, we have found the optimal thresholds for the dynamical state indicators, and fitted them to the best possible polynomial models. The results are shown in Fig. \ref{fig3}, from top to bottom, for the centre offset, virial ratio, mean radial velocity, sparsity and ellipticity thresholds.

Most of the thresholds on the assembly state indicators present a clear redshift evolution. At earlier times, the thresholds on the dynamical state indicators tend to take higher values, reflecting the fact that haloes at earlier times were more irregular or exhibited more disturbed features, even when not having experienced any relevant merging activity or growth during the last dynamical time.

The evolution of the thresholds ranges from very mild or almost nonexistent (e.g., $\Delta_r^\mathrm{thr}$, $\varepsilon^\mathrm{thr}$) to noticeable (and definitely worth taking into account; e.g., $\eta^\mathrm{thr}$, $s_{200c,500c}^\mathrm{thr}$, $\langle \widetilde{v_r} \rangle^\mathrm{thr}$). This unequivocally evidences that fixed, set thresholds on certain parameters may not be able to correctly discriminate relaxed from merging haloes through the whole evolutionary history of the objects, especially when delving into the realm of high-redshift haloes.

The thresholds can be fitted by the following equations (solid lines in Fig. \ref{fig3}, whose uncertainties are represented by the shaded regions), valid for $0 \leq z \leq 5$, where the figures in parentheses correspond to the uncertainty in the two last digits of each coefficient:

\begin{equation}
    \Delta_r^\mathrm{thr}(z) = 0.0849(13)
    \label{eq:fit_centre_offset}
\end{equation}
\begin{equation}
    \eta^\mathrm{thr}(z) = 1.3383(56) + 0.197(11) z - 0.0276(32) z^2
    \label{eq:fit_virial}
\end{equation}
\begin{equation}
    \langle \widetilde{v_r} \rangle_\mathrm{DM}^\mathrm{thr}(z) = 0.0718(22) + 0.0056(14) z
    \label{eq:fit_mean_vr}
\end{equation}
\begin{equation}
    s_{200c,500c}^\mathrm{thr}(z) = 1.491(16) + 0.064(37) z - 0.031(22) z^2 + 0.0060(35) z^3
    \label{eq:fit_sparsity}
\end{equation}
\begin{equation}
    \varepsilon^\mathrm{thr}(z) = 0.2696(27)
    \label{eq:fit_ellipticity}
\end{equation}

\begin{figure}
    \centering
    \includegraphics[width=0.8\linewidth]{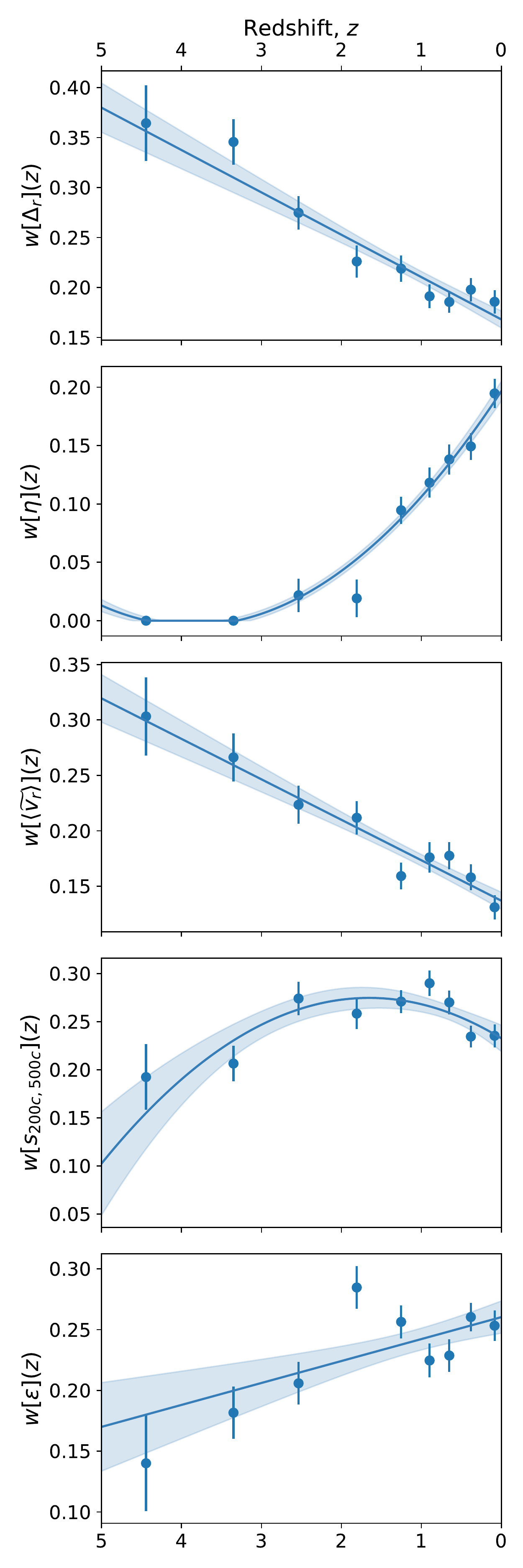}
    \caption{Redshift evolution of the weights on the dynamical state indicators. From top to bottom, the panels refer to the centre offset ($w[\Delta_r]$), virial ratio ($w[\eta]$), mean radial velocity ($w[\langle \tilde{v_r} \rangle]$), sparsity ($w[s_{200c,500c}]$), and ellipticity ($w[\varepsilon]$) weights. Dots correspond to the weights obtained within the redshift bin, with the error bars obtained by means of bootstrap resampling. Solid lines correspond to the best polynomial fits, with their (16-84)\% confidence interval as the shaded region.}
    \label{fig4}
\end{figure}

Based on the performance of each assembly state indicator in matching the fiducial classification, we fix the weights of each indicator in Eq. \ref{eq:xi_dynstate} as described in Sec. \ref{s:methods.classification.sample}. The results are summarised in Fig. \ref{fig4}, which is analogous to Fig. \ref{fig3} but this time showing the weights instead of the thresholds. Note that, if all indicators were equivalently important, $w_i=0.2 \, \forall i$. Thus, $w_i > 0.2$ ($w_i < 0.2$) implies above-average (below-average) performance for the given dynamical state indicator at the given epoch.

Interestingly, the importance of each indicator in determining the dynamical state of DM haloes varies strongly with redshift. For example, one of the most widely used indicators, the centre offset $\Delta_r$, is exceedingly effective in discriminating the disturbed haloes at high redshift, but its effectiveness declines steeply with decreasing redshift and has slightly below-average performance at $z \simeq 0$. As an example of the opposite trend, the virial ratio, $\eta$, appears to be irrelevant at high redshift ($z \gtrsim 2$), and is only useful at low redshifts ($ \lesssim 1$). This dissimilar behaviour between centre offset and virial ratio is also reported by the analyses at high redshift of \cite{Davis_2011}.

As a purely dynamical parameter, the mean radial velocity $\langle \widetilde{v_r} \rangle$ is especially relevant at high redshift, likely due to the fact that smooth (nearly radial) accretion could be more important at these stages given the relatively higher density in the surroundings of the halo. Sparsity, as well as ellipticity, are especially correlated with the fiducial dynamical state classification at more recent redshifts, although they cannot generally be neglected at any epoch. As a matter of fact, at low redshift, $\varepsilon$ is the most relevant indicator of the dynamical state of haloes.

With the same procedure as above, we have fitted the weights to polynomial functions capturing their evolution (solid lines in Fig. \ref{fig4}, whose uncertainties are represented by the shaded regions), valid for $0 \leq z \leq 5$:

\begin{equation}
    w[\Delta_r](z) \propto 0.1679(70) + 0.0423(50) z
    \label{eq:fit_centre_offset_weight}
\end{equation}
\begin{equation}
    w[\eta](z) \propto 0.1965(78) - 0.1037(60) z + 0.0134(11) z^2
    \label{eq:fit_virial_weight}
\end{equation}
\begin{equation}
    w[\langle \widetilde{v_r} \rangle_\mathrm{DM}](z) \propto 0.1370(70) + 0.0364(48) z
    \label{eq:fit_mean_vr_weight}
\end{equation}
\begin{equation}
    w[s_{200c,500c}](z) \propto 0.2327(97) + 0.051(14) z - 0.0153(38) z^2
    \label{eq:fit_sparsity_weight}
\end{equation}
\begin{equation}
    w[\varepsilon](z) \propto 0.2603(75) - 0.0181(51) z
    \label{eq:fit_ellipticity_weight}
\end{equation}

We note that, while at any epoch the data points fulfilled $\sum_{i=1}^5 w_i = 1$, this is not necessarily true for the fitting polynomials evaluated at any arbitrary redshift (although it holds to a few percents). Thus, they must be normalised by their sum before plugging them into Eq. \ref{eq:xi_dynstate}.

\begin{figure}
    \centering
    \includegraphics[width=\linewidth]{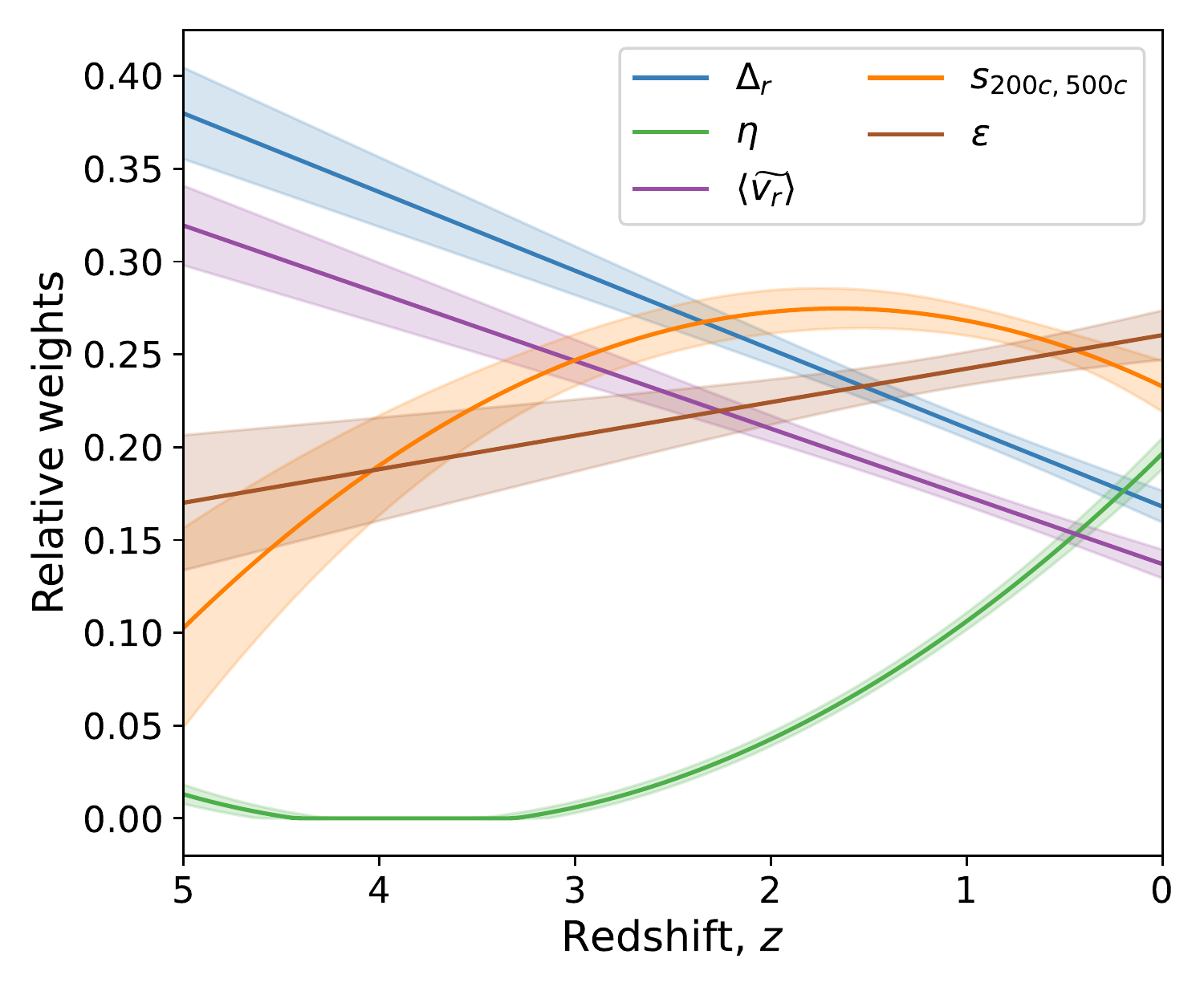}
    \caption{Overall fitted redshift evolution of the weights of the dynamical state indicators, using the complete sample.}
    \label{fig5}
\end{figure}

For better comparison of the relative importance of each of the indicators, Fig. \ref{fig5} presents the fits of the weights on each dynamical state indicator, as a function of decreasing redshift. The behaviour can be roughly summarised as:

\begin{itemize}
    \item At high redshift ($z \in [2,5]$), the centre offset provides the most insightful information about the recent assembly activity. This can be primarily complemented by the mean radial velocity (especially at $z \gtrsim 3$), or sparsity and ellipticity (at $z \lesssim 3$). The virial ratio does not seem to provide any insight on the dynamical state at high redshift.
    \item At intermediate redshifts, ($z \in [1,2]$), sparsity, ellipticity and centre offset provide similarly useful information about the dynamical state. The relevance of the virial ratio is still limited at this epoch.
    \item  At low redshifts ($z \lesssim 1$), the ellipticity of the DM halo correlates exceptionally well with the dynamical state, as well as sparsity does. While no dynamical state indicator is negligible at this stage, centre offset and virial ratio also present reasonable performances, while mean radial velocity is the least useful indicator at this time.
\end{itemize}

\subsection{Dependence on halo mass}
\label{s:results.mass_dependence}

The previous analyses have considered all haloes on an equal footing, despite their broad distribution in masses. Invoking self-similarity (see, e.g., \citealp{Navarro_2010} for a thorough analysis on the level of self-similarity of haloes), it may be argued that the same thresholds and weights could be used for all halo masses. However, the fact that many halo properties (related to the dynamical state indicators in our work) scale with mass demands to explicitly check how our results depend on the mass scale of the haloes being considered.

\begin{figure}
    \centering
    \includegraphics[width=\linewidth]{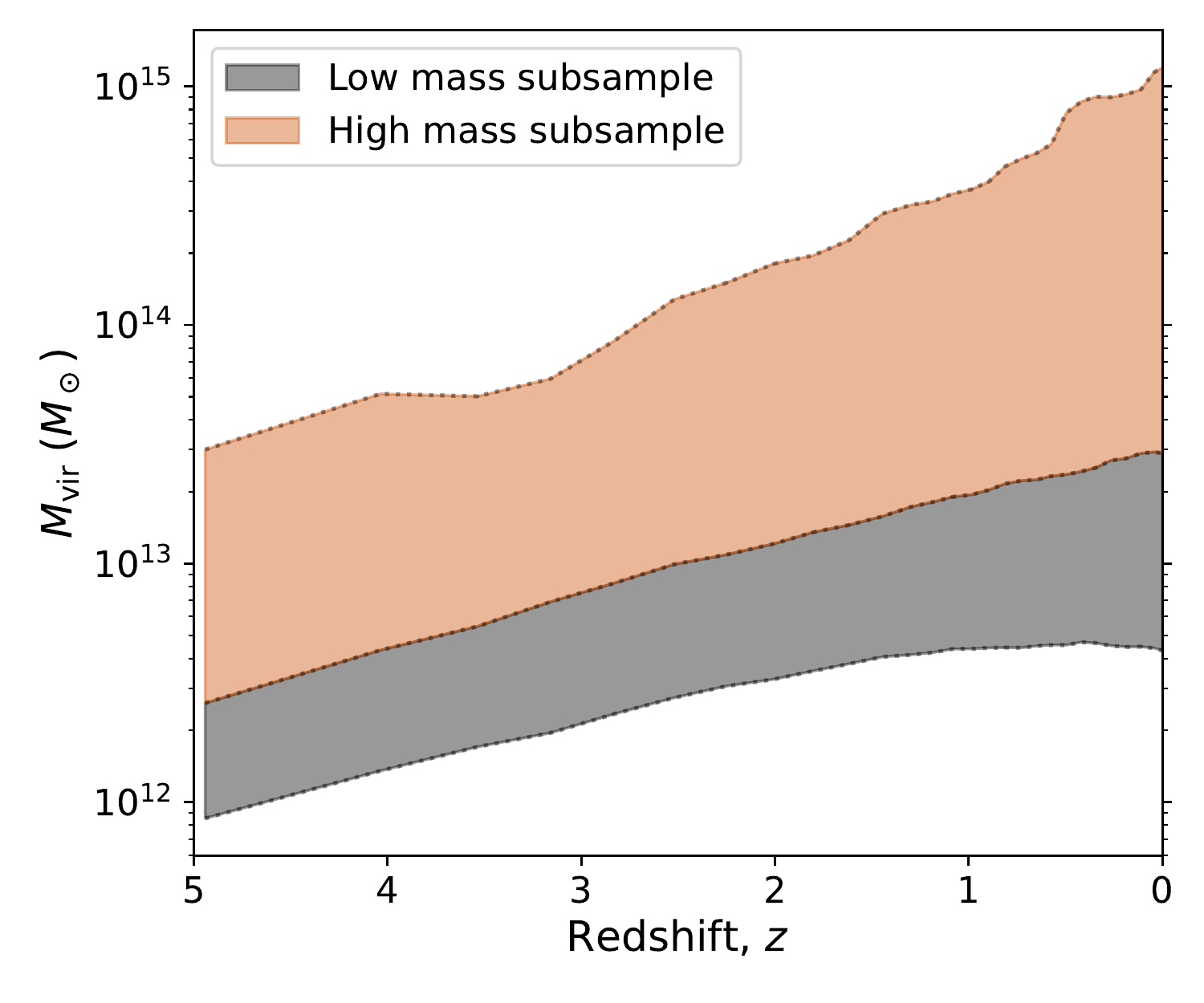}
    \caption{Definition of the mass subsamples in terms of mass, as a function of redshift. The gray (salmon) shaded regions correspond to the low-mass (high-mass) subsamples. The dotted lines mark the mass limits.}
    \label{fig6}
\end{figure}

We have split the complete sample, introduced in Sec. \ref{s:methods.classification.sample}, in two subsamples, namely a low-mass and a high-mass subsample. The high-mass subsample contains, at each redshift, the $20\%$ most massive haloes in the complete sample. This is chosen so as to contain, at redshift 0, all the haloes associated to massive groups and clusters ($M_\mathrm{DM} > 3 \times 10^{13} M_\odot$). Fig. \ref{fig6} shows the evolution of the mass limits on each subsample. Note that, therefore, our mass groups do not correspond to fixed-mass ranges, but rather to two sub-populations with a redshift-dependent mass threshold. While it would definetely be interesting to explore the dependence of our thresholds and weights with actual mass ranges, our limited statistics prevent us from this goal and we may defer this for future work.

\begin{figure}
    \centering
    \includegraphics[width=0.8\linewidth]{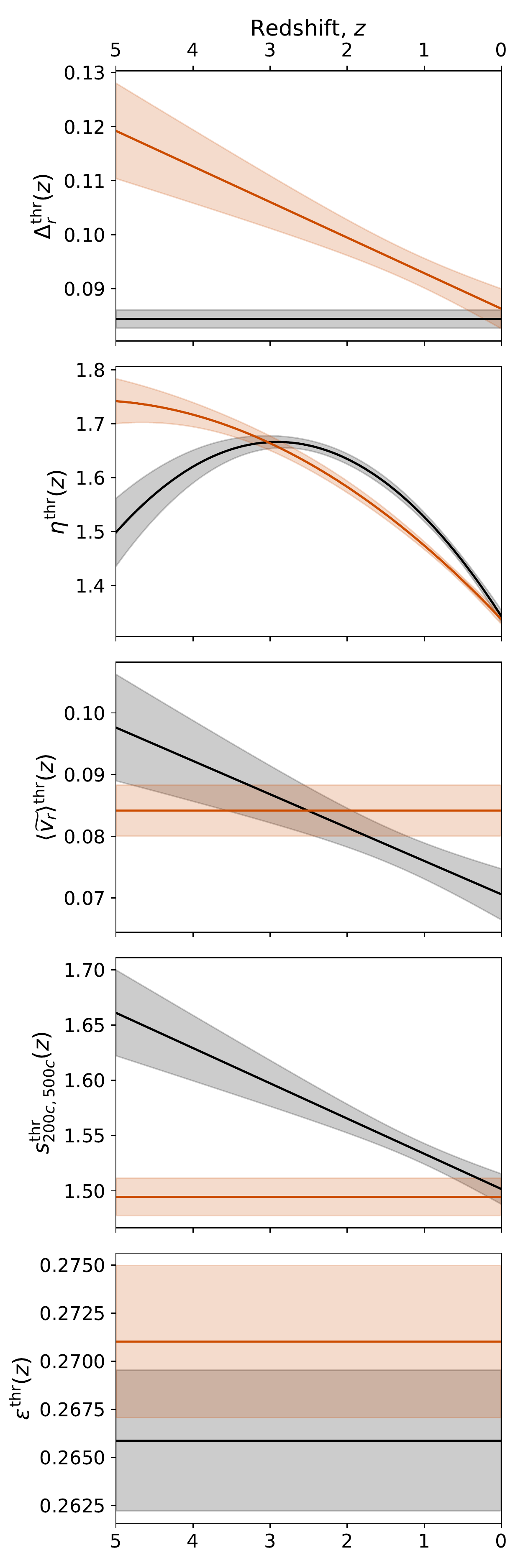}
    \caption{Mass dependence of the redshift evolution of the thresholds on the dynamical state indicators. The figure is analogous to Fig. \ref{fig3}, with each line corresponding to the fit performed with a mass subsample (the same colour coding as in Fig. \ref{fig6} is used: gray [salmon] lines correspond to the low-mass [high-mass] subsamples), and the shaded regions enclosing $1\sigma$ confidence intervals.}
    \label{fig7}
\end{figure}

Repeating the analyses above separately for each mass subsample, we can infer the mass dependence of the thresholds on the dynamical state indicators, and that of their corresponding weights in Eq. \ref{eq:xi_dynstate}. The redshift evolution of the thresholds for each mass subsample is presented in Fig. \ref{fig7}, which is analogous to Fig. \ref{fig3} but displaying only the fits for clarity. The same colour coding as in Fig. \ref{fig6} is used here.

For some indicators, such as ellipticity, there is no hint for any significant trend of the evolution of the threshold with mass, at least within the statistical uncertainties given by our sample size. That is to say, at least within the mass range considered in this work (roughly, $[10^{12} - 10^{15}] \, M_\odot$), the relaxation criteria based on this indicator can be used regardless of the scale of the objects (as customarily done with many indicators, e.g. \citealp{Power_2012}).

However, the rest of indicators of the dynamical state do present significant dependence on halo mass. In particular, the threshold on $\Delta_r$ remains constant with redshift for the low-mass subsample, while it increases linearly with increasing redshift for group and cluster-sized DM haloes. This would suggest that imposing a constant threshold on the centre offset may be too conservative and could, for massive haloes at high redshift, artificially increase in excess the number of disturbed haloes.

Regarding virial ratio, there is a minor trend with mass at intermediate and low redshifts ($z \lesssim 3$), with higher-mass haloes preferring a slightly more stringent threshold to separate dynamically relaxed and unrelaxed haloes, but this difference corresponds to a small variation on the value of the parameter ($\Delta \eta^\mathrm{thr} \sim 0.05$). The most striking difference appears at high redshift, but is not relevant since we have found that $\eta$ itself is not meaningful at high redshift (see the second panel in Fig. \ref{fig4}, as well as Fig. \ref{fig8} below).

The mass dependence of $\langle \widetilde{v_r} \rangle$ is moderate, with massive haloes preferring a constant threshold around $\langle \widetilde{v_r} \rangle \approx 0.085$ and low-mass systems displaying a decreasing trend with decreasing redshift. 

Last, the two mass subsamples present a different behaviours with respect to the threshold on halo sparsity, $s_{200c,500c}^\mathrm{thr}$. In this case, lower mass haloes require consistently larger thresholds on sparsity to discriminate relaxed and merging objects. While smaller haloes tend to be more concentrated \citep[see, for instance,][]{Dutton_2014}, the mass-dependence of sparsity is much more contained \citep{Corasaniti_2018, Corasaniti_2019}. It might be the case that, both for physical (e.g., stronger influence of the environment) and numerical (e.g., less mass resolution elements leading to more unresolved central regions) reasons, low-mass haloes present a broader distribution in sparsities (see, e.g., figure 10 in \citealp{Balmes_2014}) and thus a larger sparsity threshold is required at the low mass end.

\begin{figure*}
    \centering
    \includegraphics[width=0.66\linewidth]{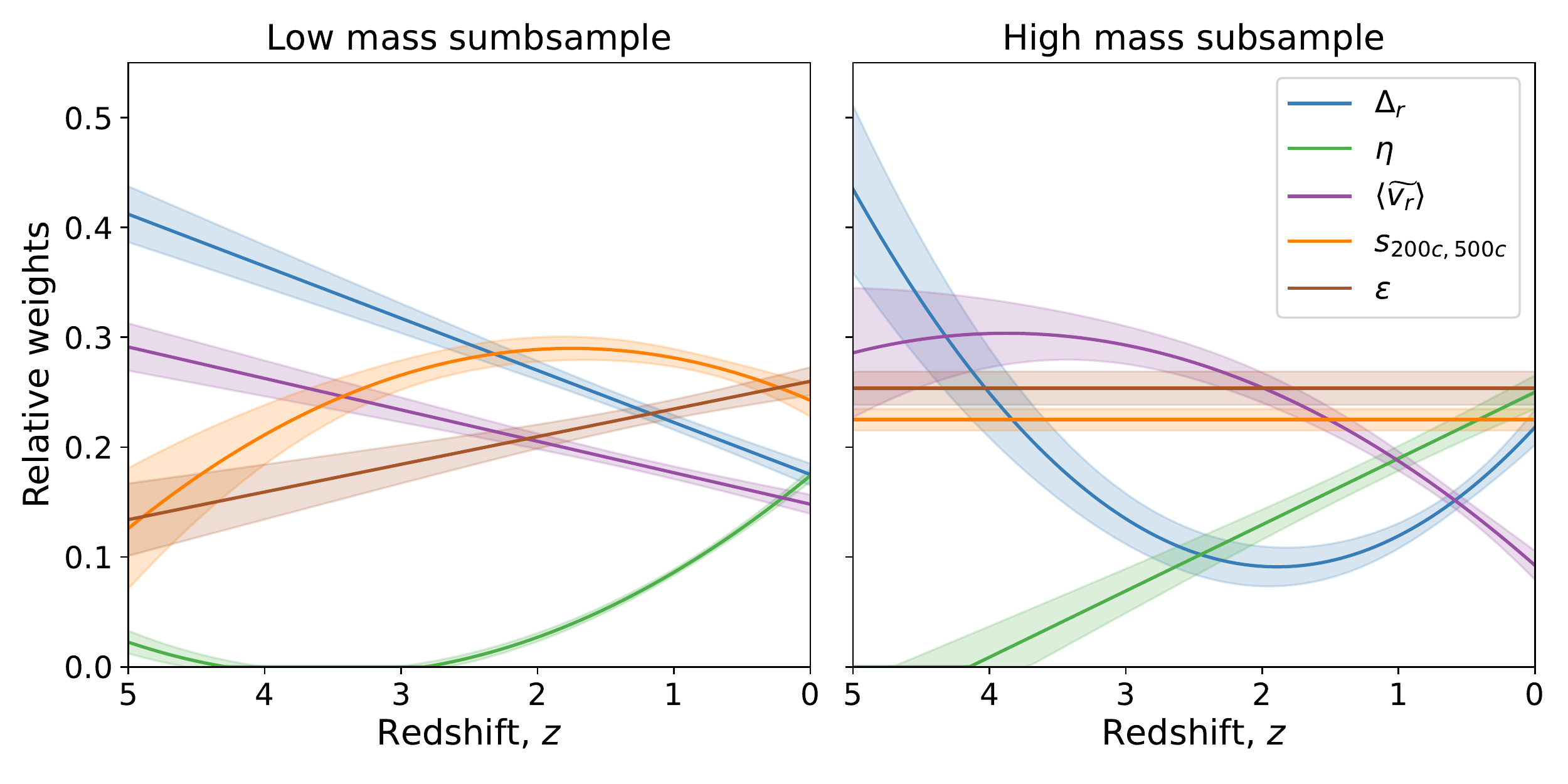}
    \caption{Mass dependence of the redshift evolution of the weights on the dynamical state indicators. Each panel is analogous to Fig. \ref{fig5} (also using the same colour coding), showing the results for each of the mass samples defined in Fig. \ref{fig6}: low-mass (left panel) and high-mass (right panel) subsamples.}
    \label{fig8}
\end{figure*}

Finally, we show in Fig. \ref{fig8} the evolution of the weights on each of the dynamical state indicators (as they appear in Eq. \ref{eq:xi_dynstate}), for each of the mass subsamples. Besides the general trends already analysed for the whole sample when Fig. \ref{fig4} was presented, several differences emerge between the high-mass and the low-mass subsamples, especially at intermediate and high redshifts. At low redshift, however, the weights are essentially compatible amongst the two subsamples, with only a small hint of virial ratio and centre offset being --comparatively-- more effective in higher-mass haloes.

At intermediate redshifts, $z \sim 2$-$3$, high-mass haloes find ellipticity and mean radial velocity to be better indicators of their dynamical state, and centre offset and sparsity comparatively worse ones, when confronted to the low-mass sample. At high redshifts, $z \sim 5$, the performance of sparsity gets more penalised for lower-mass haloes and, in these objects, centre offset can be relatively more important than in higher-mass haloes. This further highlights that, besides not being able to put fix criteria (in the sense of them not evolving with redshift) for assessing the dynamical state of dark matter haloes, they have to be carefully chosen depending on the scale of the object being studied.

We provide fits for the thresholds and weights for the high-mass (group and cluster-sized) subsample in Appendix \ref{s:appA}.

\subsection{Classification assessment}
\label{s:results.why_5}

The aim of this section is to validate to which extent the dynamical state classification introduced in this work, which only uses information at a given timestep, is capable of predicting the merging state of the DM halo. That is to say, whether we can predict the fiducial classification (Sec. \ref{s:methods.haloes.fiducial}) based on the dynamical state indicators, when confronting our method with haloes from a different simulation (corresponding to different resolution, gravity solver, etc.).

We use public simulation data from the suite \texttt{CAMELS} \citep{Villaescusa-Navarro_2021, Villaescusa-Navarro_2022}, which contains over 4000 simulations of $25 h^{-1} \, \mathrm{Mpc}$ cubic, periodic volumes run with different physics, cosmological and astrophysical parameters, and numerical codes. In particular, we have analysed the haloes in the \texttt{IllustrisTNG-DM} \texttt{CV-0} simulation, which corresponds to a DM-only simulation run with \texttt{Arepo} \citep{Springel_2010, Weinberger_2020}. \texttt{Arepo} implements a Tree+Particle-Mesh approach \citep{Bagla_2002} for solving the evolution of DM, thus providing a high dynamical range even though the number of particles ($N_\mathrm{part}=256^3$) is rather small. The \texttt{CV-0} simulation corresponds to a background cosmology with $h=0.6711$, $\Omega_m = 0.3$, $n_s=0.9624$, and $\sigma_8=0.8$, the initial conditions having been set up at $z_\mathrm{ini}=127$.

We have extracted the halo catalogues and merger trees with \texttt{ASOHF} \citep{Valles_2022} by following the exact same procedure described in Sec. \ref{s:methods.haloes}, and determined the dynamical state indicators (Sec. \ref{s:methods.indicators}). For our analyses, we have considered the 30 most massive haloes at each time, which corresponds to a similar mass limit as in the main analysis (cf. Fig. \ref{fig1}). Out of these 30 haloes, we have dropped the ones that we are not able to trace back in time for at least one dynamical time (which is most usually none or one halo, at the considered epochs).

\begin{figure*}
    \centering
    \includegraphics[width=\textwidth]{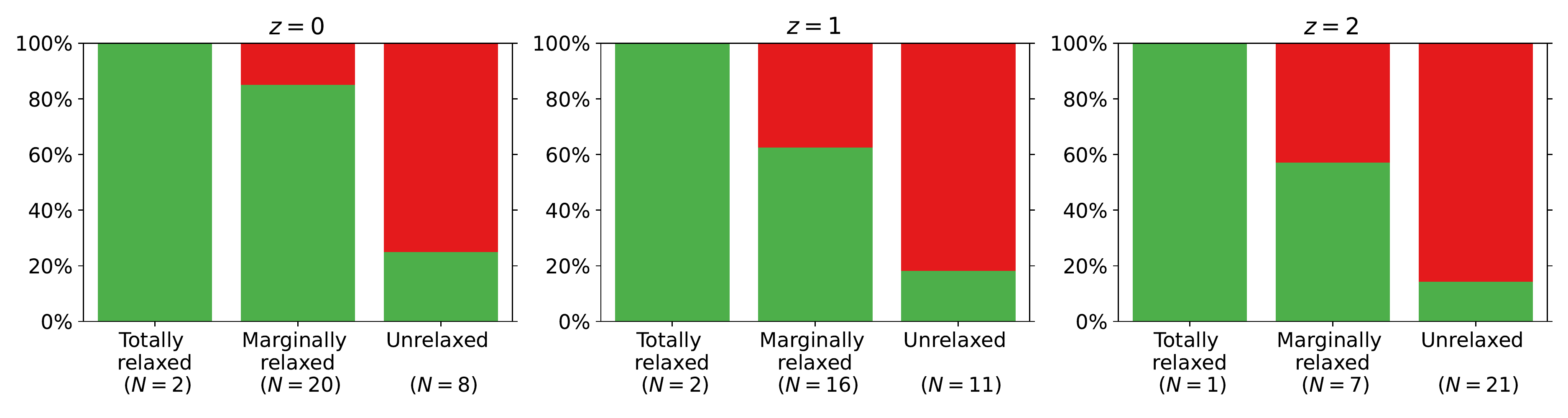}
    \caption{{Classification summary at redshifts $z=0$ (left panel), $z=1$ (middle panel), and $z=2$ (right panel). Within each panel, each of the columns corresponds to one of the possible classifications (totally relaxed, if $X_i < X_i^\mathrm{thr} \; \forall i$; marginally relaxed, if the previous condition fails but $\chi \geq 1$; or unrelaxed otherwise). Within each bar, the green portion represents the fraction of the haloes which have not suffered any mergers nor strong accretion, while the red portion corresponds to the fraction of haloes having suffered mergers or strong accretion (i.e., the colour encodes the fiducial classification). Below each column, $N$ indicates the number of objects falling into each category.}}
    \label{fig9}
\end{figure*}

In Fig. \ref{fig9}, we assess the performance of our classification scheme at three cosmological epochs ($z=0, \, 1, \, \text{and} \; 2$, for the panels left to right) by computing, at each time, the fraction of haloes in each class (totally relaxed, marginally relaxed and unrelaxed) which have recently suffered mergers or strong accretion (red), or has undergone a quiet evolution (green). We note that, when evaluating the dynamical state criteria, at any given $z$ we only apply the indicators with weight $w_i(z) > 0.05$. Otherwise, we consider the given dynamical state indicator as not meaningful at that particular epoch. While this particular threshold is arbitrary, it is a sensible choice and the results do not depend strongly on variations around this value. According to Fig. \ref{fig5}, this only removes the virial ratio, $\eta$, at $z \gtrsim 1.9$.

The totally relaxed subsample is, naturally, the smallest one, since it is defined rather conservatively as the set of haloes simultaneously fulfilling all five relaxation criteria. It typically contains $\sim 10\%$ of the haloes (slightly lower in this case; nevertheless, the statistics are small). Within this test, all haloes being classified as totally relaxed have not suffered any major (minor) merger within one (half) $\tau_\mathrm{dyn}$ or built up more than half of their mass in the last dynamical time, thus proving to be a selection of the relaxed sample with high specificity.

The marginally relaxed sample is the most numerous at low redshifts ($z \lesssim 1$) and mostly contains objects which have not experienced any relevant merging or accretion activity, although the fraction of objects having experienced it  increases with increasing redshift (from $\sim 15\%$ at $z\simeq 0$ to $\sim 40\%$ at $z\simeq 2$). The unrelaxed subsample, which is especially numerous at high redshift when merger rates are higher (see, e.g., \citealp{Wetzel_2009}), contains mostly merging objects, although a small fraction ($10-25\%$) of objects not experiencing mergers or accretion seem to fall into this category. This may happen because a halo appears to be disturbed, even when not merging or accreting intensely, due to environmental effects (e.g., strong tidal field due to the presence of another nearby massive halo, for instance in a pre-merger state), or even numerical effects (mainly associated to low resolution). Naturally, it may also be the case that unrelaxedness after a major merger is last for longer than $1 \tau_\mathrm{dyn}(z)$, since the fiducial classification in Sec. \ref{s:methods.haloes.fiducial} was only a rough estimation.

\begin{figure*}
    \centering
    \includegraphics[width=\textwidth]{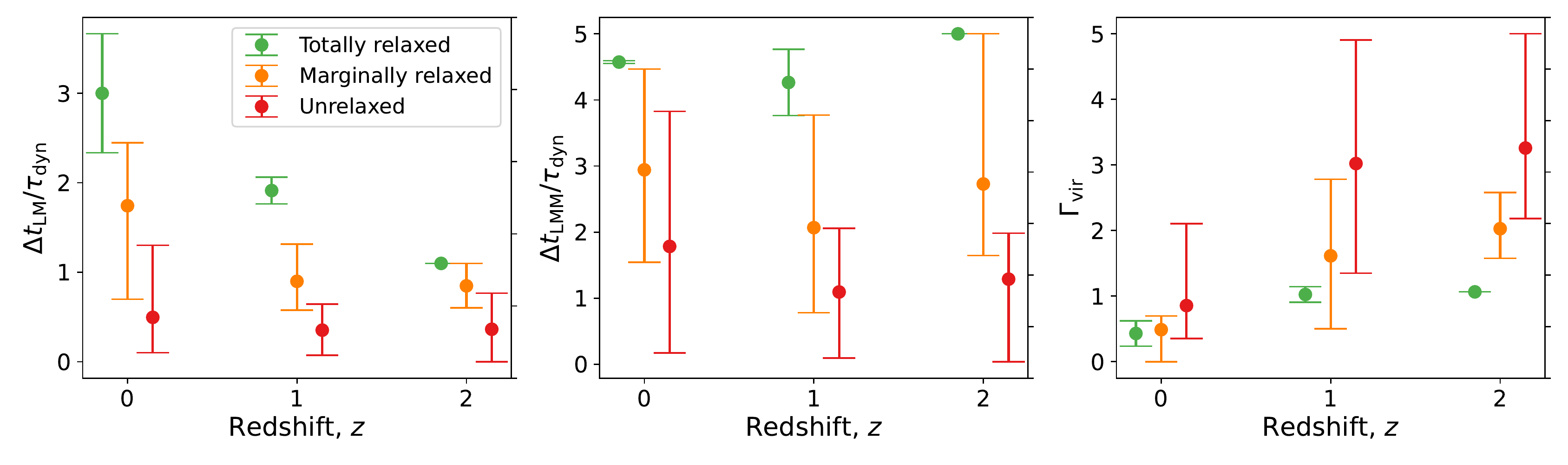}
    \caption{Trends of the evolutionary properties of the haloes according to their dynamical state classification. In each panel, red, orange and green dots correspond, respectively, to the unrelaxed, marginally relaxed and totally relaxed subsamples. At each redshift (encoded in the horizontal axis) the dot and the error bars represent, respectively, the mean and the $1\sigma$ dispersion of the distribution of the given variable within the subsample. The left panel presents the time since the last merger (either major or minor) in units of the dynamical time ($\Delta t_\mathrm{LM} / \tau_\mathrm{dyn}$), the central panel corresponds to the time since the last major merger ($\Delta t_\mathrm{LMM} / \tau_\mathrm{dyn}$), and the right panel shows the accretion rate $\Gamma_\mathrm{vir}$.}
    \label{fig10}
\end{figure*}

Expanding upon the previous figure, in Fig. \ref{fig10} we focus on some quantities tied to the merger and accretion history of the haloes. In particular, the left panel represents, at three redshifts ($z=0,\,1,\, \text{and}\; 2$, respectively, from left to right) and for the three subsamples, the distribution of the time since the last merger (either major or minor) in units of the dynamical time. Haloes classified as unrelaxed have usually suffered some merger recently while, on the other hand, the totally relaxed sample has typically not experienced any merging activity since several dynamical times ago. A similar situation is seen for the major mergers (middle panel), although naturally not all unrelaxed haloes have suffered a major merger (the unrelaxedness can be due to one or several minor mergers, or smooth accretion, as well). However, the same trend holds, with major mergers having occurred a longer time ago as we move from unrelaxed to marginally relaxed, and to totally relaxed haloes.

Finally, the third panel presents, in a similar way, the distribution of accretion rates $\Gamma_\mathrm{vir}$, which are computed according to the prescription of \cite{Diemer_2014},

\begin{equation}
    \Gamma_\mathrm{vir} = \frac{\Delta \log M_\mathrm{vir}}{\Delta \log a}
    \label{eq:MAR}
\end{equation}

\noindent with $a$ being the scale factor of the Universe, and the increments computed over a dynamical time following its definition in Eq. \ref{eq:dyntime_integral}. The figure shows, in line with the previous results, an increasing trend of the accretion rate when moving from the relaxed to the more disturbed subsamples, within a wide redshift interval. This reflects how the dynamical state classification presented in this work, which only uses information at a fixed time, can offer insight on the temporal evolution of the systems over the last dynamical time.

\subsubsection{Does a smaller set of indicators provide similar insight?}

\begin{table*}
    \centering
    \caption{Classification properties using only one or a combination of dynamical state indicators, exemplified at $z=1$. Each row corresponds to one dynamical indicator or combination, and for each classification class we give the number of objects falling into the class ($N$, and the percentage with respect to the total) and the fraction of them which is unrelaxed according to the fiducial classification ($f_\mathrm{merging}$). When only one indicator is used, there is no intermediate (\textit{marginally relaxed}) class. The first block corresponds to the individual indicators involved in this work. The second block contains several combinations widely used in the literature, with the fitted thresholds in this work (for $f_\mathrm{sub}$, not involved in this work, we use $f_\mathrm{sub}^\mathrm{thr}=0.1$). The last row corresponds to the complete method introduced here, using the five indicators (therefore, these results are the same shown in the central panel of Fig. \ref{fig9}).}
    \begin{tabular}{c|cc|cc|cc}
         & \multicolumn{2}{c|}{Totally relaxed} & \multicolumn{2}{c|}{Marginally relaxed} & \multicolumn{2}{c}{Unrelaxed}  \\
         Indicator(s)  & $N$ & $f_\mathrm{merging}$ & $N$ & $f_\mathrm{merging}$ & $N$ & $f_\mathrm{merging}$ \\ \hline

        $\Delta_r$ & 20 (69$\%$)& 0.40 & -- & -- & 9 (31$\%$)& 0.78 \\
        $\eta$ & 25 (86$\%$)& 0.48 & -- & -- & 4 (14$\%$)& 0.75 \\
        $\langle \tilde v_r \rangle$ & 21 (72$\%$)& 0.43 & -- & -- & 8 (28$\%$)& 0.75 \\
        $s_{200c,500c}$ & 17 (59$\%$)& 0.47 & -- & -- & 12 (41$\%$)& 0.58 \\
        $\epsilon$ & 10 (34$\%$)& 0.50 & -- & -- & 19 (66$\%$)& 0.53 \\ \hline
        $\Delta_r$ \& $\eta$ & 19 (66$\%$)& 0.42 & 7 (24$\%$)& 0.57 & 3 (10$\%$)& 1.00 \\
        $f_\mathrm{sub}$ & 23 (79$\%$)& 0.48 & -- & -- & 6 (21$\%$)& 0.67 \\
        $\eta$ \& $f_\mathrm{sub}$ & 18 (62$\%$)& 0.44 & 7 (24$\%$)& 0.43 & 4 (14$\%$)& 1.00 \\
        $\Delta_r$, \, $\eta$ \& $f_\mathrm{sub}$ & 17 (59$\%$)& 0.47 & 3 (10$\%$)& 0.33 & 9 (31$\%$)& 0.67 \\ \hline\hline
        Full set of indicators & 2 (7$\%$) & 0.00 & 16 (55$\%$) & 0.38 & 11 (38$\%$) & 0.82 \\ \hline  \hline  
    \end{tabular}
    \label{tab2}
\end{table*}

Lastly, it might be interesting to assess whether a single dynamical state indicator, or a combination of them, is capable of providing a similarly accurate classification; i.e., to motivate why it is important to involve a high number of indicators. This is briefly exemplified in Table \ref{tab2}, where we show the classification summary for each indicator (or combination). In particular, for each classification class we give the number of haloes falling into this class (and its percentage with respect to the total), and the fraction of them which is unrelaxed according to the fiducial classification ($f_\mathrm{merging}$). Ideally, this fraction would be 0 for the totally relaxed class and 1 for the unrelaxed class. Naturally, when using only one indicator, there is no \textit{marginally relaxed} or intermediate category, since the value of the dynamical state indicator can only be above or below the threshold. In the case of using two indicators, we have defined the marginally relaxed sample as the set of haloes fulfilling only one of the two relaxedness conditions, as it is often done in the literature (e.g. \citealp{Biffi_2016, Planelles_2017}).

Generally speaking, involving only one dynamical state indicator leads to far poorer results, since the relaxed sample gets often contaminated (around $\sim 40\%$) by haloes which have suffered mergers. Likewise, the unrelaxed sample may end up containing a high fraction of haloes undergoing quiescent evolution for some indicators (e.g., $\epsilon$; although the particular results have to be considered carefully due to the reduced statistics).

Interestingly, when using a combination of virial ratio and centre offset, which is a common option in the literature (e.g., \citealp{Power_2012}), we are still not able to pick out all merging haloes with these criteria and even the totally relaxed subsample gets contaminated with $\sim 40\%$ of merging haloes. Other common options in the literature are the combination of mass ratio and centre offset \citep{DeLuca_2021}, or centre offset, virial ratio and mass ratio \citep{Cui_2017, Haggar_2020}. We have also tested these combinations, taking $f_\mathrm{sub}^\mathrm{thr}=0.1$ from the aforementioned references, since we have not involved this indicator in our previous analyses. In these cases, the results are similar to the $\Delta_r$ \& $\eta$ combination. This highlights the necessity of involving and combining as many indicators of the dynamical state as possible. When using the full set of indicators derived in this work, the totally relaxed subsample is rather small, due to its conservative definition. However, even our marginally relaxed subsample is purer (contains a smaller fraction of merging/accreting haloes) than the totally relaxed sample of the previous combinations, proving to provide a robust splitting of haloes according to their dynamical state.

\section{Discussion and conclusions}
\label{s:conclusions}

To fully exploit the capabilities of ongoing surveys (e.g., \texttt{eROSITA}; \citealp{Ghiradini_2022}), and in the advent of upcoming instruments over the electromagnetic spectrum (from X-ray, e.g., \texttt{ATHENA}, \citealp{Nandra_2013}; to radio, e.g. \texttt{SKA}, \citealp{AcostaPulido_2015}; going through the optical, e.g. \texttt{EUCLID}, \citealp{Sartoris_2016, Euclid_2019}), which will provide samples of galaxies and galaxy clusters unprecedented in size and depth, it remains crucial to provide reliable indicators of the dynamical state (which in turns is a fast proxy of the --recent-- evolution of the system). As a first step towards that aim, using $N$-Body+hydrodynamics simulations, in this work we have systematically analysed how to best combine a series of quantities which can be measured from simulation data at a given time in order to be able to detect the presence of mergers and/or ongoing strong accretion. 

As a result, we have built an algorithm that combines a series of different indicators of the dynamical state of a DM halo (namely, its centre offset $\Delta_r$, the virial ratio $\eta$, the mean radial velocity $\langle \widetilde {v_r} \rangle$, the sparsity $s_{200c,500c}$, and the ellipticity $\varepsilon$) in order to classify haloes within three classes. The \textit{totally relaxed} haloes, comprising the objects simultaneously fulfilling all relaxedness conditions (which are redshift-dependent, in general), is a conservatively defined subsample which, therefore, only contains around $\sim 10\%$ of the haloes at a given time. Haloes where some relaxedness condition may fail, but are remarkably relaxed according to the rest of indicators may be categorised in the \textit{marginally relaxed} class, using a criterion similar to \cite{Haggar_2020}, but allowing different indicators to have different (redshift-dependent) weights, which are tuned based on the performance of each indicator on telling relaxed and unrelaxed haloes apart. Thus, we defined a \textit{relaxedness parametre} ($\chi$), which tells \textit{marginally relaxed} ($\chi\geq 1$) and \textit{unrelaxed} ($\chi<1$) apart. The fits for the redshift dependence of the thresholds and weights are given in Eqns. \ref{eq:fit_centre_offset}-\ref{eq:fit_ellipticity_weight}, while equivalent results for \textit{massive} haloes are provided in App. \ref{s:appA}.

Furthermore, we have confronted our classification scheme against an independent DM-only simulation from the \texttt{CAMELS} suite \citep{Villaescusa-Navarro_2021, Villaescusa-Navarro_2022}, corresponding to different input physics, initial conditions and numerical solvers. Using it, we find that our algorithm performs a clean splitting of relaxed and unrelaxed haloes across a wide cosmic time interval, and that this classification improves upon the usage of any single indicator or some widely used combinations ($\Delta_r$ \& $\eta$; $f_\mathrm{sub}$ \& $\eta$; or $\Delta_r$, $\eta$ \& $f_\mathrm{sub}$).

As a qualitative summary of the main highlights of the classification scheme, we can mention:

\begin{itemize}
    \item Placing fix thresholds (which do not evolve with redshift) is generally undesirable. While some indicators do not show strong evolution of their optimal thresholds with redshift (e.g., ellipticity, centre offset), others do (e.g., sparsity, mean radial velocity; all tending to increase with redshift). This has important consequences, since it implies that classification schemes for the dynamical state of haloes that are set at $z=0$ cannot be directly used at high redshifts. 
    \begin{itemize}
        \item At high halo mass (see the precise definition of the high-mass subsample in Fig. \ref{fig6}), however, the results are slightly changed: in particular, the redshift dependence of the thresholds on $\langle \widetilde{v_r}\rangle$ and $s_{200c,500c}$ is not significant anymore, while the classification based on the $\Delta_r$ benefits from an increasing trend with increasing redshift. This warns us that the classification cannot be universal, and that haloes on different mass scales may need slightly modified criteria.
    \end{itemize}
    \item At low redshift ($z \lesssim 1$), even though all indicators offer insight into the merging state of the halo, it is sparsity and ellipticity of the DM halo the ones which provide the most valuable information, well beyond other, more widely used indicators such as centre offset or virial ratio. Nevertheless, the fact that all relative weights are not very dissimilar at this epoch (see Fig. \ref{fig5}) means that the classification scheme can importantly benefit from combining as many indicators as possible.
    \begin{itemize}
        \item The difference in weights amongst the different observables (except $\langle \widetilde{v_r}\rangle$) is importantly reduced when looking at the high mass sample (right panel in Fig. \ref{fig8}), reinforcing that, for group- and cluster-sized haloes at low redshift, it may be important to combine all indicators suggested in this work.
    \end{itemize}    
    \item At high redshifts ($z \gtrsim 3$), $\eta$ becomes irrelevant for the determination of the assembly state of the halo, while centre offset and mean radial velocity become, by far, the dominant indicators. 
    \begin{itemize}
        \item Again, the differences are lower for the high-mass subsample, but the prevalence of $\Delta_r$ and $\langle \widetilde{v_r}\rangle$ still holds.
    \end{itemize}  
\end{itemize}

In this work, we have focused on the determination of the assembly state of DM haloes using the full information contained in a snapshot of a numerical simulation. The motivation for this is two-fold. On the one hand, it is important to devise efficient methods to classify large samples of simulated haloes, especially given the ever-growing trend of simulations, both in size and resolution (see, e.g., \citealp{Angulo_2022}, their table 1), made possible by the increasing computational power available. On the other hand, it serves as a first step, which can be further connected to observations using projected data or, more realistically, mock multiwavelength observations (e.g., \citealp{Planelles_2018}).

Much of the information comprised in the dynamical state indicators we involve in this work can be lost, or at least hindered, when moving from the 3-dimensional description to the 2-dimensional observed data. The first, most natural consequence is the effect of projection on any geometrical indicator, such as the centre offset, ellipticity or the mean radial velocity. For the case of centre offset and ellipticity, the measured values will only be a lower limit, with the actual 3-dimensional value depending on the inclination between the direction of the offset, or the plane containing the major and minor axis, with the line of sight.

Regarding the mean radial velocity, which is especially important for determining the dynamical state at high redshift, besides the difficulty induced by projection (only velocities along the line of sight, and distances on the plane of the sky, can be measured), future kinetic Sunyaev-Zel'dovich (kSZ) observations could be able to provide some constraints on proper velocities of the intra-cluster medium (ICM; see, for instance, the estimates of \citealp{Baldi_2018} about the kSZ effect due to the coherent rotation of the ICM), even for high-redshift objects since the SZ effect is essentially distance-independent (e.g., \citealp{Voit_2005}). Even though the dynamics of the ICM, especially in the inner regions of haloes, may differ significantly from those of the DM halo, probing the velocity field of the diffuse gas in haloes could supply useful insight onto the dynamical state of haloes at high redshift.

Lastly, sparsity may be a suitable option for observations, given its good performance shown across the whole redshift span considered here (especially, for high-mass haloes). However, care must be taken when using this quantity: here, we have defined sparsity from the DM masses obtained from the full, 3-dimensional information. However, in observations, masses can be obtained from several methods (e.g., hydrostatic, lensing, caustic masses), and biases amongst them are non-negligible (see, for instance, \citealp{Lovisari_2020}). Moreover, mass biases tend to correlate with the merging state (\citealp{Bennett_2022}; cf. \citealp{Gianfagna_2022}) and, while the quotient of two masses at different apertures derived from the same method may cancel out part of these biases, the fact that the bias itself depends on the aperture and the large object-to-object scatter still make the interpretation non-trivial and deserve further attention themselves.

This work provides a motivated definition of a scheme for classifying DM haloes according to their dynamical status, based on simple properties which can be readily extracted from the outputs of typical halo finders. Future work will need to deal with the connection of these dynamical and morphological properties of the DM halo with the baryonic component, as well as the application to observations, in order to being able to extract the largest possible amount of information about the assembly state of haloes from future observational campaigns.

\section*{Acknowledgements}

We gratefully thank the anonymous referee for their valuable feedback, which has helped us to improve the quality of this manuscript. This work has been supported by the Agencia Estatal de Investigación Española (AEI; grant PID2019-107427GB-C33), by the Ministerio de Ciencia e Innovación (MICIN) en el marco del Plan de Recuperación, Transformación y Resiliencia del Gobierno de España through the project ASFAE/2022/001 and by the Generalitat Valenciana (grant PROMETEO/2019/071). DV acknowledges support from Universitat de València through an Atracció de Talent fellowship, and gratefully thanks the hospitality of the Dipartimento di Fisica e Astronomia of the Università di Bologna, where part of this work was done during a research stay funded by Universitat de València.  We also thank F. Vazza and A. Ragagnin for fruitful scientific conversations. Simulations have been carried out with the supercomputer Lluís Vives at the Servei d’Informàtica of the Universitat de València. This research has made use of the following open-source packages: \texttt{NumPy} \citep{Numpy}, \texttt{SciPy} \citep{Scipy}, \texttt{matplotlib} \citep{Matplotlib}, \texttt{statsmodels} \citep{Statsmodels}, \texttt{scikit-learn} \citep{Scikit-learn}, and \texttt{Colossus} \citep{Diemer_2018}.

\section*{Data Availability}

The data underlying this article will be shared upon reasonable request to the corresponding author.



\bibliographystyle{mnras}
\bibliography{mnras_dynstate} 

\begin{thebibliography}{}
\makeatletter
\relax
\def\mn@urlcharsother{\let\do\@makeother \do\$\do\&\do\#\do\^\do\_\do\%\do\~}
\def\mn@doi{\begingroup\mn@urlcharsother \@ifnextchar [ {\mn@doi@}
  {\mn@doi@[]}}
\def\mn@doi@[#1]#2{\def\@tempa{#1}\ifx\@tempa\@empty \href
  {http://dx.doi.org/#2} {doi:#2}\else \href {http://dx.doi.org/#2} {#1}\fi
  \endgroup}
\def\mn@eprint#1#2{\mn@eprint@#1:#2::\@nil}
\def\mn@eprint@arXiv#1{\href {http://arxiv.org/abs/#1} {{\tt arXiv:#1}}}
\def\mn@eprint@dblp#1{\href {http://dblp.uni-trier.de/rec/bibtex/#1.xml}
  {dblp:#1}}
\def\mn@eprint@#1:#2:#3:#4\@nil{\def\@tempa {#1}\def\@tempb {#2}\def\@tempc
  {#3}\ifx \@tempc \@empty \let \@tempc \@tempb \let \@tempb \@tempa \fi \ifx
  \@tempb \@empty \def\@tempb {arXiv}\fi \@ifundefined
  {mn@eprint@\@tempb}{\@tempb:\@tempc}{\expandafter \expandafter \csname
  mn@eprint@\@tempb\endcsname \expandafter{\@tempc}}}

\bibitem[\protect\citeauthoryear{{Acosta-Pulido} et~al.,}{{Acosta-Pulido}
  et~al.}{2015}]{AcostaPulido_2015}
{Acosta-Pulido} J.~A.,  et~al., 2015, arXiv e-prints, \href
  {https://ui.adsabs.harvard.edu/abs/2015arXiv150603474A} {p. arXiv:1506.03474}

\bibitem[\protect\citeauthoryear{{Angelinelli}, {Vazza}, {Giocoli}, {Ettori},
  {Jones}, {Brunetti}, {Br{\"u}ggen}  \& {Eckert}}{{Angelinelli}
  et~al.}{2020}]{Angelinelli_2020}
{Angelinelli} M.,  {Vazza} F.,  {Giocoli} C.,  {Ettori} S.,  {Jones} T.~W.,
  {Brunetti} G.,  {Br{\"u}ggen} M.,   {Eckert} D.,  2020, \mn@doi [\mnras]
  {10.1093/mnras/staa975}, \href
  {https://ui.adsabs.harvard.edu/abs/2020MNRAS.495..864A} {495, 864}

\bibitem[\protect\citeauthoryear{{Angulo} \& {Hahn}}{{Angulo} \&
  {Hahn}}{2022}]{Angulo_2022}
{Angulo} R.~E.,  {Hahn} O.,  2022, \mn@doi [Living Reviews in Computational
  Astrophysics] {10.1007/s41115-021-00013-z}, \href
  {https://ui.adsabs.harvard.edu/abs/2022LRCA....8....1A} {8, 1}

\bibitem[\protect\citeauthoryear{{Bagla}}{{Bagla}}{2002}]{Bagla_2002}
{Bagla} J.~S.,  2002, \mn@doi [Journal of Astrophysics and Astronomy]
  {10.1007/BF02702282}, \href
  {https://ui.adsabs.harvard.edu/abs/2002JApA...23..185B} {23, 185}

\bibitem[\protect\citeauthoryear{{Baldi}, {De Petris}, {Sembolini}, {Yepes},
  {Lamagna}  \& {Rasia}}{{Baldi} et~al.}{2017}]{Baldi_2017}
{Baldi} A.~S.,  {De Petris} M.,  {Sembolini} F.,  {Yepes} G.,  {Lamagna} L.,
  {Rasia} E.,  2017, \mn@doi [\mnras] {10.1093/mnras/stw2858}, \href
  {https://ui.adsabs.harvard.edu/abs/2017MNRAS.465.2584B} {465, 2584}

\bibitem[\protect\citeauthoryear{{Baldi}, {De Petris}, {Sembolini}, {Yepes},
  {Cui}  \& {Lamagna}}{{Baldi} et~al.}{2018}]{Baldi_2018}
{Baldi} A.~S.,  {De Petris} M.,  {Sembolini} F.,  {Yepes} G.,  {Cui} W.,
  {Lamagna} L.,  2018, \mn@doi [\mnras] {10.1093/mnras/sty1722}, \href
  {https://ui.adsabs.harvard.edu/abs/2018MNRAS.479.4028B} {479, 4028}

\bibitem[\protect\citeauthoryear{{Balm{\`e}s}, {Rasera}, {Corasaniti}  \&
  {Alimi}}{{Balm{\`e}s} et~al.}{2014}]{Balmes_2014}
{Balm{\`e}s} I.,  {Rasera} Y.,  {Corasaniti} P.~S.,   {Alimi} J.~M.,  2014,
  \mn@doi [\mnras] {10.1093/mnras/stt2050}, \href
  {https://ui.adsabs.harvard.edu/abs/2014MNRAS.437.2328B} {437, 2328}

\bibitem[\protect\citeauthoryear{{Bennett} \& {Sijacki}}{{Bennett} \&
  {Sijacki}}{2022}]{Bennett_2022}
{Bennett} J.~S.,  {Sijacki} D.,  2022, \mn@doi [\mnras]
  {10.1093/mnras/stac1216}, \href
  {https://ui.adsabs.harvard.edu/abs/2022MNRAS.514..313B} {514, 313}

\bibitem[\protect\citeauthoryear{{Berger} \& {Colella}}{{Berger} \&
  {Colella}}{1989}]{Berger_Colella_1989}
{Berger} M.~J.,  {Colella} P.,  1989, \mn@doi [Journal of Computational
  Physics] {10.1016/0021-9991(89)90035-1}, \href
  {https://ui.adsabs.harvard.edu/abs/1989JCoPh..82...64B} {82, 64}

\bibitem[\protect\citeauthoryear{{Biffi} et~al.,}{{Biffi}
  et~al.}{2016}]{Biffi_2016}
{Biffi} V.,  et~al., 2016, \mn@doi [\apj] {10.3847/0004-637X/827/2/112}, \href
  {https://ui.adsabs.harvard.edu/abs/2016ApJ...827..112B} {827, 112}

\bibitem[\protect\citeauthoryear{{Bryan} \& {Norman}}{{Bryan} \&
  {Norman}}{1998}]{BryanNorman_98}
{Bryan} G.~L.,  {Norman} M.~L.,  1998, \mn@doi [\apj] {10.1086/305262}, \href
  {https://ui.adsabs.harvard.edu/abs/1998ApJ...495...80B} {495, 80}

\bibitem[\protect\citeauthoryear{{Buote} \& {Tsai}}{{Buote} \&
  {Tsai}}{1995}]{Buote_1995}
{Buote} D.~A.,  {Tsai} J.~C.,  1995, \mn@doi [\apj] {10.1086/176326}, \href
  {https://ui.adsabs.harvard.edu/abs/1995ApJ...452..522B} {452, 522}

\bibitem[\protect\citeauthoryear{{Capalbo}, {De Petris}, {De Luca}, {Cui},
  {Yepes}, {Knebe}  \& {Rasia}}{{Capalbo} et~al.}{2021}]{Capalbo_2021}
{Capalbo} V.,  {De Petris} M.,  {De Luca} F.,  {Cui} W.,  {Yepes} G.,  {Knebe}
  A.,   {Rasia} E.,  2021, \mn@doi [\mnras] {10.1093/mnras/staa3900}, \href
  {https://ui.adsabs.harvard.edu/abs/2021MNRAS.503.6155C} {503, 6155}

\bibitem[\protect\citeauthoryear{{Cerini}, {Cappelluti}  \&
  {Natarajan}}{{Cerini} et~al.}{2022}]{Cerini_2022}
{Cerini} G.,  {Cappelluti} N.,   {Natarajan} P.,  2022, arXiv e-prints, \href
  {https://ui.adsabs.harvard.edu/abs/2022arXiv220906831C} {p. arXiv:2209.06831}

\bibitem[\protect\citeauthoryear{{Chandrasekhar}}{{Chandrasekhar}}{1961}]{Chandrasekhar_1961}
{Chandrasekhar} S.,  1961, {Hydrodynamic and hydromagnetic stability}

\bibitem[\protect\citeauthoryear{{Chen}, {Wang}, {Mo}  \& {Shi}}{{Chen}
  et~al.}{2016}]{Chen_2016}
{Chen} S.,  {Wang} H.,  {Mo} H.~J.,   {Shi} J.,  2016, \mn@doi [\apj]
  {10.3847/0004-637X/825/1/49}, \href
  {https://ui.adsabs.harvard.edu/abs/2016ApJ...825...49C} {825, 49}

\bibitem[\protect\citeauthoryear{{Chen}, {Avestruz}, {Kravtsov}, {Lau}  \&
  {Nagai}}{{Chen} et~al.}{2019}]{Chen_2019}
{Chen} H.,  {Avestruz} C.,  {Kravtsov} A.~V.,  {Lau} E.~T.,   {Nagai} D.,
  2019, \mn@doi [\mnras] {10.1093/mnras/stz2776}, \href
  {https://ui.adsabs.harvard.edu/abs/2019MNRAS.490.2380C} {490, 2380}

\bibitem[\protect\citeauthoryear{{Cialone}, {De Petris}, {Sembolini}, {Yepes},
  {Baldi}  \& {Rasia}}{{Cialone} et~al.}{2018}]{Cialone_2018}
{Cialone} G.,  {De Petris} M.,  {Sembolini} F.,  {Yepes} G.,  {Baldi} A.~S.,
  {Rasia} E.,  2018, \mn@doi [\mnras] {10.1093/mnras/sty621}, \href
  {https://ui.adsabs.harvard.edu/abs/2018MNRAS.477..139C} {477, 139}

\bibitem[\protect\citeauthoryear{{Cole} \& {Lacey}}{{Cole} \&
  {Lacey}}{1996}]{Cole_1996}
{Cole} S.,  {Lacey} C.,  1996, \mn@doi [\mnras] {10.1093/mnras/281.2.716},
  \href {https://ui.adsabs.harvard.edu/abs/1996MNRAS.281..716C} {281, 716}

\bibitem[\protect\citeauthoryear{{Corasaniti} \& {Rasera}}{{Corasaniti} \&
  {Rasera}}{2019}]{Corasaniti_2019}
{Corasaniti} P.~S.,  {Rasera} Y.,  2019, \mn@doi [\mnras]
  {10.1093/mnras/stz1579}, \href
  {https://ui.adsabs.harvard.edu/abs/2019MNRAS.487.4382C} {487, 4382}

\bibitem[\protect\citeauthoryear{{Corasaniti}, {Ettori}, {Rasera}, {Sereno},
  {Amodeo}, {Breton}, {Ghirardini}  \& {Eckert}}{{Corasaniti}
  et~al.}{2018}]{Corasaniti_2018}
{Corasaniti} P.~S.,  {Ettori} S.,  {Rasera} Y.,  {Sereno} M.,  {Amodeo} S.,
  {Breton} M.~A.,  {Ghirardini} V.,   {Eckert} D.,  2018, \mn@doi [\apj]
  {10.3847/1538-4357/aaccdf}, \href
  {https://ui.adsabs.harvard.edu/abs/2018ApJ...862...40C} {862, 40}

\bibitem[\protect\citeauthoryear{{Crone}, {Evrard}  \& {Richstone}}{{Crone}
  et~al.}{1996}]{Crone_1996}
{Crone} M.~M.,  {Evrard} A.~E.,   {Richstone} D.~O.,  1996, \mn@doi [\apj]
  {10.1086/177626}, \href
  {https://ui.adsabs.harvard.edu/abs/1996ApJ...467..489C} {467, 489}

\bibitem[\protect\citeauthoryear{{Cui} et~al.,}{{Cui} et~al.}{2016}]{Cui_2016}
{Cui} W.,  et~al., 2016, \mn@doi [\mnras] {10.1093/mnras/stv2839}, \href
  {https://ui.adsabs.harvard.edu/abs/2016MNRAS.456.2566C} {456, 2566}

\bibitem[\protect\citeauthoryear{{Cui}, {Power}, {Borgani}, {Knebe}, {Lewis},
  {Murante}  \& {Poole}}{{Cui} et~al.}{2017}]{Cui_2017}
{Cui} W.,  {Power} C.,  {Borgani} S.,  {Knebe} A.,  {Lewis} G.~F.,  {Murante}
  G.,   {Poole} G.~B.,  2017, \mn@doi [\mnras] {10.1093/mnras/stw2567}, \href
  {https://ui.adsabs.harvard.edu/abs/2017MNRAS.464.2502C} {464, 2502}

\bibitem[\protect\citeauthoryear{{D'Onghia} \& {Navarro}}{{D'Onghia} \&
  {Navarro}}{2007}]{dOnghia_2007}
{D'Onghia} E.,  {Navarro} J.~F.,  2007, \mn@doi [\mnras]
  {10.1111/j.1745-3933.2007.00348.x}, \href
  {https://ui.adsabs.harvard.edu/abs/2007MNRAS.380L..58D} {380, L58}

\bibitem[\protect\citeauthoryear{{Davis}, {D'Aloisio}  \& {Natarajan}}{{Davis}
  et~al.}{2011}]{Davis_2011}
{Davis} A.~J.,  {D'Aloisio} A.,   {Natarajan} P.,  2011, \mn@doi [\mnras]
  {10.1111/j.1365-2966.2011.19026.x}, \href
  {https://ui.adsabs.harvard.edu/abs/2011MNRAS.416..242D} {416, 242}

\bibitem[\protect\citeauthoryear{{De Luca}, {De Petris}, {Yepes}, {Cui},
  {Knebe}  \& {Rasia}}{{De Luca} et~al.}{2021}]{DeLuca_2021}
{De Luca} F.,  {De Petris} M.,  {Yepes} G.,  {Cui} W.,  {Knebe} A.,   {Rasia}
  E.,  2021, \mn@doi [\mnras] {10.1093/mnras/stab1073}, \href
  {https://ui.adsabs.harvard.edu/abs/2021MNRAS.504.5383D} {504, 5383}

\bibitem[\protect\citeauthoryear{{Diemer}}{{Diemer}}{2018}]{Diemer_2018}
{Diemer} B.,  2018, \mn@doi [\apjs] {10.3847/1538-4365/aaee8c}, \href
  {https://ui.adsabs.harvard.edu/abs/2018ApJS..239...35D} {239, 35}

\bibitem[\protect\citeauthoryear{{Diemer} \& {Kravtsov}}{{Diemer} \&
  {Kravtsov}}{2014}]{Diemer_2014}
{Diemer} B.,  {Kravtsov} A.~V.,  2014, \mn@doi [\apj]
  {10.1088/0004-637X/789/1/1}, \href
  {https://ui.adsabs.harvard.edu/abs/2014ApJ...789....1D} {789, 1}

\bibitem[\protect\citeauthoryear{{Dutton} \& {Macci{\`o}}}{{Dutton} \&
  {Macci{\`o}}}{2014}]{Dutton_2014}
{Dutton} A.~A.,  {Macci{\`o}} A.~V.,  2014, \mn@doi [\mnras]
  {10.1093/mnras/stu742}, \href
  {https://ui.adsabs.harvard.edu/abs/2014MNRAS.441.3359D} {441, 3359}

\bibitem[\protect\citeauthoryear{Efron}{Efron}{1979}]{Efron_1979}
Efron B.,  1979, \mn@doi [The Annals of Statistics] {10.1214/aos/1176344552},
  7, 1

\bibitem[\protect\citeauthoryear{{Eisenstein} \& {Hu}}{{Eisenstein} \&
  {Hu}}{1998}]{Eisenstein_1998}
{Eisenstein} D.~J.,  {Hu} W.,  1998, \mn@doi [\apj] {10.1086/305424}, \href
  {https://ui.adsabs.harvard.edu/abs/1998ApJ...496..605E} {496, 605}

\bibitem[\protect\citeauthoryear{{Eke}, {Cole}  \& {Frenk}}{{Eke}
  et~al.}{1996}]{Eke_1996}
{Eke} V.~R.,  {Cole} S.,   {Frenk} C.~S.,  1996, \mn@doi [\mnras]
  {10.1093/mnras/282.1.263}, \href
  {https://ui.adsabs.harvard.edu/abs/1996MNRAS.282..263E} {282, 263}

\bibitem[\protect\citeauthoryear{{Euclid Collaboration} et~al.,}{{Euclid
  Collaboration} et~al.}{2019}]{Euclid_2019}
{Euclid Collaboration} et~al., 2019, \mn@doi [\aap]
  {10.1051/0004-6361/201935088}, \href
  {https://ui.adsabs.harvard.edu/abs/2019A&A...627A..23E} {627, A23}

\bibitem[\protect\citeauthoryear{Fawcett}{Fawcett}{2006}]{Fawcett_2006}
Fawcett T.,  2006, \mn@doi [Pattern Recognition Letters]
  {https://doi.org/10.1016/j.patrec.2005.10.010}, 27, 861

\bibitem[\protect\citeauthoryear{{Franx}, {Illingworth}  \& {de Zeeuw}}{{Franx}
  et~al.}{1991}]{Franx_1991}
{Franx} M.,  {Illingworth} G.,   {de Zeeuw} T.,  1991, \mn@doi [\apj]
  {10.1086/170769}, \href
  {https://ui.adsabs.harvard.edu/abs/1991ApJ...383..112F} {383, 112}

\bibitem[\protect\citeauthoryear{{Frenk}, {White}, {Davis}  \&
  {Efstathiou}}{{Frenk} et~al.}{1988}]{Frenk_1988}
{Frenk} C.~S.,  {White} S. D.~M.,  {Davis} M.,   {Efstathiou} G.,  1988,
  \mn@doi [\apj] {10.1086/166213}, \href
  {https://ui.adsabs.harvard.edu/abs/1988ApJ...327..507F} {327, 507}

\bibitem[\protect\citeauthoryear{{Ghirardini} et~al.,}{{Ghirardini}
  et~al.}{2022}]{Ghiradini_2022}
{Ghirardini} V.,  et~al., 2022, \mn@doi [\aap] {10.1051/0004-6361/202141639},
  \href {https://ui.adsabs.harvard.edu/abs/2022A&A...661A..12G} {661, A12}

\bibitem[\protect\citeauthoryear{{Gianfagna}, {Rasia}, {Cui}, {De Petris}  \&
  {Yepes}}{{Gianfagna} et~al.}{2022}]{Gianfagna_2022}
{Gianfagna} G.,  {Rasia} E.,  {Cui} W.,  {De Petris} M.,   {Yepes} G.,  2022,
  in mm Universe @ NIKA2 - Observing the mm Universe with the NIKA2 Camera. p.
  00020 (\mn@eprint {arXiv} {2111.01903}),
  \mn@doi{10.1051/epjconf/202225700020}

\bibitem[\protect\citeauthoryear{{Gott} \& {Rees}}{{Gott} \&
  {Rees}}{1975}]{Gott_1975}
{Gott} J.~R. I.,  {Rees} M.~J.,  1975, \aap, \href
  {https://ui.adsabs.harvard.edu/abs/1975A&A....45..365G} {45, 365}

\bibitem[\protect\citeauthoryear{{Gouin}, {Bonnaire}  \& {Aghanim}}{{Gouin}
  et~al.}{2021}]{Gouin_2021}
{Gouin} C.,  {Bonnaire} T.,   {Aghanim} N.,  2021, \mn@doi [\aap]
  {10.1051/0004-6361/202140327}, \href
  {https://ui.adsabs.harvard.edu/abs/2021A&A...651A..56G} {651, A56}

\bibitem[\protect\citeauthoryear{{Gouin}, {Gallo}  \& {Aghanim}}{{Gouin}
  et~al.}{2022}]{Gouin_2022}
{Gouin} C.,  {Gallo} S.,   {Aghanim} N.,  2022, arXiv e-prints, \href
  {https://ui.adsabs.harvard.edu/abs/2022arXiv220100593G} {p. arXiv:2201.00593}

\bibitem[\protect\citeauthoryear{{Haggar}, {Gray}, {Pearce}, {Knebe}, {Cui},
  {Mostoghiu}  \& {Yepes}}{{Haggar} et~al.}{2020}]{Haggar_2020}
{Haggar} R.,  {Gray} M.~E.,  {Pearce} F.~R.,  {Knebe} A.,  {Cui} W.,
  {Mostoghiu} R.,   {Yepes} G.,  2020, \mn@doi [\mnras]
  {10.1093/mnras/staa273}, \href
  {https://ui.adsabs.harvard.edu/abs/2020MNRAS.492.6074H} {492, 6074}

\bibitem[\protect\citeauthoryear{Harris et~al.,}{Harris et~al.}{2020}]{Numpy}
Harris C.~R.,  et~al., 2020, \mn@doi [Nature] {10.1038/s41586-020-2649-2}, 585,
  357

\bibitem[\protect\citeauthoryear{{Hockney} \& {Eastwood}}{{Hockney} \&
  {Eastwood}}{1988}]{Hockney_Eastwood_1988}
{Hockney} R.~W.,  {Eastwood} J.~W.,  1988, {Computer simulation using
  particles}

\bibitem[\protect\citeauthoryear{Hunter}{Hunter}{2007}]{Matplotlib}
Hunter J.~D.,  2007, \mn@doi [Computing in Science \& Engineering]
  {10.1109/MCSE.2007.55}, 9, 90

\bibitem[\protect\citeauthoryear{{Jiang} \& {van den Bosch}}{{Jiang} \& {van
  den Bosch}}{2016}]{Jiang_2016}
{Jiang} F.,  {van den Bosch} F.~C.,  2016, \mn@doi [\mnras]
  {10.1093/mnras/stw439}, \href
  {https://ui.adsabs.harvard.edu/abs/2016MNRAS.458.2848J} {458, 2848}

\bibitem[\protect\citeauthoryear{{Kimmig}, {Remus}, {Dolag}  \&
  {Biffi}}{{Kimmig} et~al.}{2022}]{Kimmig_2022}
{Kimmig} L.~C.,  {Remus} R.-S.,  {Dolag} K.,   {Biffi} V.,  2022, arXiv
  e-prints, \href {https://ui.adsabs.harvard.edu/abs/2022arXiv220909916K} {p.
  arXiv:2209.09916}

\bibitem[\protect\citeauthoryear{{Knebe} \& {Power}}{{Knebe} \&
  {Power}}{2008}]{Knebe_2008}
{Knebe} A.,  {Power} C.,  2008, \mn@doi [\apj] {10.1086/586702}, \href
  {https://ui.adsabs.harvard.edu/abs/2008ApJ...678..621K} {678, 621}

\bibitem[\protect\citeauthoryear{{Knebe} \& {Wie{\ss}ner}}{{Knebe} \&
  {Wie{\ss}ner}}{2006}]{Knebe_2006}
{Knebe} A.,  {Wie{\ss}ner} V.,  2006, \mn@doi [\pasa] {10.1071/AS06013}, \href
  {https://ui.adsabs.harvard.edu/abs/2006PASA...23..125K} {23, 125}

\bibitem[\protect\citeauthoryear{{Knebe} et~al.,}{{Knebe}
  et~al.}{2011}]{Knebe_2011}
{Knebe} A.,  et~al., 2011, \mn@doi [\mnras] {10.1111/j.1365-2966.2011.18858.x},
  \href {https://ui.adsabs.harvard.edu/abs/2011MNRAS.415.2293K} {415, 2293}

\bibitem[\protect\citeauthoryear{{Kravtsov} \& {Borgani}}{{Kravtsov} \&
  {Borgani}}{2012}]{Kravtsov_2012}
{Kravtsov} A.~V.,  {Borgani} S.,  2012, \mn@doi [\araa]
  {10.1146/annurev-astro-081811-125502}, \href
  {https://ui.adsabs.harvard.edu/abs/2012ARA&A..50..353K} {50, 353}

\bibitem[\protect\citeauthoryear{{Kuchner} et~al.,}{{Kuchner}
  et~al.}{2020}]{Kuchner_2020}
{Kuchner} U.,  et~al., 2020, \mn@doi [\mnras] {10.1093/mnras/staa1083}, \href
  {https://ui.adsabs.harvard.edu/abs/2020MNRAS.494.5473K} {494, 5473}

\bibitem[\protect\citeauthoryear{{Kuchner} et~al.,}{{Kuchner}
  et~al.}{2022}]{Kuchner_2022}
{Kuchner} U.,  et~al., 2022, \mn@doi [\mnras] {10.1093/mnras/stab3419}, \href
  {https://ui.adsabs.harvard.edu/abs/2022MNRAS.510..581K} {510, 581}

\bibitem[\protect\citeauthoryear{{Lau}, {Hearin}, {Nagai}  \&
  {Cappelluti}}{{Lau} et~al.}{2021}]{Lau_2021}
{Lau} E.~T.,  {Hearin} A.~P.,  {Nagai} D.,   {Cappelluti} N.,  2021, \mn@doi
  [\mnras] {10.1093/mnras/staa3313}, \href
  {https://ui.adsabs.harvard.edu/abs/2021MNRAS.500.1029L} {500, 1029}

\bibitem[\protect\citeauthoryear{{Lovisari}, {Ettori}, {Sereno},
  {Schellenberger}, {Forman}, {Andrade-Santos}  \& {Jones}}{{Lovisari}
  et~al.}{2020}]{Lovisari_2020}
{Lovisari} L.,  {Ettori} S.,  {Sereno} M.,  {Schellenberger} G.,  {Forman}
  W.~R.,  {Andrade-Santos} F.,   {Jones} C.,  2020, \mn@doi [\aap]
  {10.1051/0004-6361/202038718}, \href
  {https://ui.adsabs.harvard.edu/abs/2020A&A...644A..78L} {644, A78}

\bibitem[\protect\citeauthoryear{{Macci{\`o}}, {Dutton}, {van den Bosch},
  {Moore}, {Potter}  \& {Stadel}}{{Macci{\`o}} et~al.}{2007}]{Maccio_2007}
{Macci{\`o}} A.~V.,  {Dutton} A.~A.,  {van den Bosch} F.~C.,  {Moore} B.,
  {Potter} D.,   {Stadel} J.,  2007, \mn@doi [\mnras]
  {10.1111/j.1365-2966.2007.11720.x}, \href
  {https://ui.adsabs.harvard.edu/abs/2007MNRAS.378...55M} {378, 55}

\bibitem[\protect\citeauthoryear{{Mohr}, {Fabricant}  \& {Geller}}{{Mohr}
  et~al.}{1993}]{Mohr_1993}
{Mohr} J.~J.,  {Fabricant} D.~G.,   {Geller} M.~J.,  1993, \mn@doi [\apj]
  {10.1086/173019}, \href
  {https://ui.adsabs.harvard.edu/abs/1993ApJ...413..492M} {413, 492}

\bibitem[\protect\citeauthoryear{{Nandra} et~al.,}{{Nandra}
  et~al.}{2013}]{Nandra_2013}
{Nandra} K.,  et~al., 2013, arXiv e-prints, \href
  {https://ui.adsabs.harvard.edu/abs/2013arXiv1306.2307N} {p. arXiv:1306.2307}

\bibitem[\protect\citeauthoryear{{Navarro}, {Frenk}  \& {White}}{{Navarro}
  et~al.}{1997}]{Navarro_1997}
{Navarro} J.~F.,  {Frenk} C.~S.,   {White} S. D.~M.,  1997, \mn@doi [\apj]
  {10.1086/304888}, \href
  {https://ui.adsabs.harvard.edu/abs/1997ApJ...490..493N} {490, 493}

\bibitem[\protect\citeauthoryear{{Navarro} et~al.,}{{Navarro}
  et~al.}{2010}]{Navarro_2010}
{Navarro} J.~F.,  et~al., 2010, \mn@doi [\mnras]
  {10.1111/j.1365-2966.2009.15878.x}, \href
  {https://ui.adsabs.harvard.edu/abs/2010MNRAS.402...21N} {402, 21}

\bibitem[\protect\citeauthoryear{{Nelson}, {Lau}, {Nagai}, {Rudd}  \&
  {Yu}}{{Nelson} et~al.}{2014}]{Nelson_2014}
{Nelson} K.,  {Lau} E.~T.,  {Nagai} D.,  {Rudd} D.~H.,   {Yu} L.,  2014,
  \mn@doi [\apj] {10.1088/0004-637X/782/2/107}, \href
  {https://ui.adsabs.harvard.edu/abs/2014ApJ...782..107N} {782, 107}

\bibitem[\protect\citeauthoryear{{Neto} et~al.,}{{Neto}
  et~al.}{2007}]{Neto_2007}
{Neto} A.~F.,  et~al., 2007, \mn@doi [\mnras]
  {10.1111/j.1365-2966.2007.12381.x}, \href
  {https://ui.adsabs.harvard.edu/abs/2007MNRAS.381.1450N} {381, 1450}

\bibitem[\protect\citeauthoryear{Pedregosa et~al.,}{Pedregosa
  et~al.}{2011}]{Scikit-learn}
Pedregosa F.,  et~al., 2011, Journal of Machine Learning Research, 12, 2825

\bibitem[\protect\citeauthoryear{{Planck Collaboration} et~al.,}{{Planck
  Collaboration} et~al.}{2020}]{Planck_2020}
{Planck Collaboration} et~al., 2020, \mn@doi [\aap]
  {10.1051/0004-6361/201833910}, \href
  {https://ui.adsabs.harvard.edu/abs/2020A&A...641A...6P} {641, A6}

\bibitem[\protect\citeauthoryear{{Planelles} \& {Quilis}}{{Planelles} \&
  {Quilis}}{2009}]{Planelles_2009}
{Planelles} S.,  {Quilis} V.,  2009, \mn@doi [\mnras]
  {10.1111/j.1365-2966.2009.15290.x}, \href
  {https://ui.adsabs.harvard.edu/abs/2009MNRAS.399..410P} {399, 410}

\bibitem[\protect\citeauthoryear{{Planelles} \& {Quilis}}{{Planelles} \&
  {Quilis}}{2010}]{Planelles_2010}
{Planelles} S.,  {Quilis} V.,  2010, \mn@doi [\aap]
  {10.1051/0004-6361/201014214}, \href
  {https://ui.adsabs.harvard.edu/abs/2010A&A...519A..94P} {519, A94}

\bibitem[\protect\citeauthoryear{{Planelles}, {Schleicher}  \&
  {Bykov}}{{Planelles} et~al.}{2015}]{Planelles_2015}
{Planelles} S.,  {Schleicher} D.~R.~G.,   {Bykov} A.~M.,  2015, \mn@doi [\ssr]
  {10.1007/s11214-014-0045-7}, \href
  {https://ui.adsabs.harvard.edu/abs/2015SSRv..188...93P} {188, 93}

\bibitem[\protect\citeauthoryear{{Planelles} et~al.,}{{Planelles}
  et~al.}{2017}]{Planelles_2017}
{Planelles} S.,  et~al., 2017, \mn@doi [\mnras] {10.1093/mnras/stx318}, \href
  {https://ui.adsabs.harvard.edu/abs/2017MNRAS.467.3827P} {467, 3827}

\bibitem[\protect\citeauthoryear{{Planelles}, {Mimica}, {Quilis}  \&
  {Cuesta-Mart{\'\i}nez}}{{Planelles} et~al.}{2018}]{Planelles_2018}
{Planelles} S.,  {Mimica} P.,  {Quilis} V.,   {Cuesta-Mart{\'\i}nez} C.,  2018,
  \mn@doi [\mnras] {10.1093/mnras/sty527}, \href
  {https://ui.adsabs.harvard.edu/abs/2018MNRAS.476.4629P} {476, 4629}

\bibitem[\protect\citeauthoryear{{Poole}, {Fardal}, {Babul}, {McCarthy},
  {Quinn}  \& {Wadsley}}{{Poole} et~al.}{2006}]{Poole_2006}
{Poole} G.~B.,  {Fardal} M.~A.,  {Babul} A.,  {McCarthy} I.~G.,  {Quinn} T.,
  {Wadsley} J.,  2006, \mn@doi [\mnras] {10.1111/j.1365-2966.2006.10916.x},
  \href {https://ui.adsabs.harvard.edu/abs/2006MNRAS.373..881P} {373, 881}

\bibitem[\protect\citeauthoryear{{Power}, {Knebe}  \& {Knollmann}}{{Power}
  et~al.}{2012}]{Power_2012}
{Power} C.,  {Knebe} A.,   {Knollmann} S.~R.,  2012, \mn@doi [\mnras]
  {10.1111/j.1365-2966.2011.19820.x}, \href
  {https://ui.adsabs.harvard.edu/abs/2012MNRAS.419.1576P} {419, 1576}

\bibitem[\protect\citeauthoryear{{Press} \& {Schechter}}{{Press} \&
  {Schechter}}{1974}]{Press_1974}
{Press} W.~H.,  {Schechter} P.,  1974, \mn@doi [\apj] {10.1086/152650}, \href
  {https://ui.adsabs.harvard.edu/abs/1974ApJ...187..425P} {187, 425}

\bibitem[\protect\citeauthoryear{{Quilis}}{{Quilis}}{2004}]{Quilis_2004}
{Quilis} V.,  2004, \mn@doi [\mnras] {10.1111/j.1365-2966.2004.08040.x}, \href
  {https://ui.adsabs.harvard.edu/abs/2004MNRAS.352.1426Q} {352, 1426}

\bibitem[\protect\citeauthoryear{{Quilis}, {Mart{\'\i}}  \&
  {Planelles}}{{Quilis} et~al.}{2020}]{Quilis_2020}
{Quilis} V.,  {Mart{\'\i}} J.-M.,   {Planelles} S.,  2020, \mn@doi [\mnras]
  {10.1093/mnras/staa877}, \href
  {https://ui.adsabs.harvard.edu/abs/2020MNRAS.494.2706Q} {494, 2706}

\bibitem[\protect\citeauthoryear{{Rasia}, {Meneghetti}  \& {Ettori}}{{Rasia}
  et~al.}{2013}]{Rasia_2013}
{Rasia} E.,  {Meneghetti} M.,   {Ettori} S.,  2013, \mn@doi [The Astronomical
  Review] {10.1080/21672857.2013.11519713}, \href
  {https://ui.adsabs.harvard.edu/abs/2013AstRv...8a..40R} {8, 40}

\bibitem[\protect\citeauthoryear{{Richardson} \& {Corasaniti}}{{Richardson} \&
  {Corasaniti}}{2022}]{Richardson_2022}
{Richardson} T.~R.~G.,  {Corasaniti} P.~S.,  2022, \mn@doi [\mnras]
  {10.1093/mnras/stac1241}, \href
  {https://ui.adsabs.harvard.edu/abs/2022MNRAS.513.4951R} {513, 4951}

\bibitem[\protect\citeauthoryear{{Rossetti} et~al.,}{{Rossetti}
  et~al.}{2016}]{Rosetti_2016}
{Rossetti} M.,  et~al., 2016, \mn@doi [\mnras] {10.1093/mnras/stw265}, \href
  {https://ui.adsabs.harvard.edu/abs/2016MNRAS.457.4515R} {457, 4515}

\bibitem[\protect\citeauthoryear{{Sartoris} et~al.,}{{Sartoris}
  et~al.}{2016}]{Sartoris_2016}
{Sartoris} B.,  et~al., 2016, \mn@doi [\mnras] {10.1093/mnras/stw630}, \href
  {https://ui.adsabs.harvard.edu/abs/2016MNRAS.459.1764S} {459, 1764}

\bibitem[\protect\citeauthoryear{Seabold \& Perktold}{Seabold \&
  Perktold}{2010}]{Statsmodels}
Seabold S.,  Perktold J.,  2010, in 9th Python in Science Conference.

\bibitem[\protect\citeauthoryear{{Seppi} et~al.,}{{Seppi}
  et~al.}{2021}]{Seppi_2021}
{Seppi} R.,  et~al., 2021, \mn@doi [\aap] {10.1051/0004-6361/202039123}, \href
  {https://ui.adsabs.harvard.edu/abs/2021A&A...652A.155S} {652, A155}

\bibitem[\protect\citeauthoryear{{Shaw}, {Weller}, {Ostriker}  \&
  {Bode}}{{Shaw} et~al.}{2006}]{Shaw_2006}
{Shaw} L.~D.,  {Weller} J.,  {Ostriker} J.~P.,   {Bode} P.,  2006, \mn@doi
  [\apj] {10.1086/505016}, \href
  {https://ui.adsabs.harvard.edu/abs/2006ApJ...646..815S} {646, 815}

\bibitem[\protect\citeauthoryear{{Simonte}, {Vazza}, {Brighenti},
  {Br{\"u}ggen}, {Jones}  \& {Angelinelli}}{{Simonte}
  et~al.}{2022}]{Simonte_2022}
{Simonte} M.,  {Vazza} F.,  {Brighenti} F.,  {Br{\"u}ggen} M.,  {Jones} T.~W.,
   {Angelinelli} M.,  2022, \mn@doi [\aap] {10.1051/0004-6361/202141703}, \href
  {https://ui.adsabs.harvard.edu/abs/2022A&A...658A.149S} {658, A149}

\bibitem[\protect\citeauthoryear{{Springel}}{{Springel}}{2010}]{Springel_2010}
{Springel} V.,  2010, \mn@doi [\mnras] {10.1111/j.1365-2966.2009.15715.x},
  \href {https://ui.adsabs.harvard.edu/abs/2010MNRAS.401..791S} {401, 791}

\bibitem[\protect\citeauthoryear{{Valdarnini}}{{Valdarnini}}{2019}]{Valdarnini_2019}
{Valdarnini} R.,  2019, \mn@doi [\apj] {10.3847/1538-4357/ab0964}, \href
  {https://ui.adsabs.harvard.edu/abs/2019ApJ...874...42V} {874, 42}

\bibitem[\protect\citeauthoryear{{Vall{\'e}s-P{\'e}rez}, {Planelles}  \&
  {Quilis}}{{Vall{\'e}s-P{\'e}rez} et~al.}{2020}]{Valles_2020}
{Vall{\'e}s-P{\'e}rez} D.,  {Planelles} S.,   {Quilis} V.,  2020, \mn@doi
  [\mnras] {10.1093/mnras/staa3035}, \href
  {https://ui.adsabs.harvard.edu/abs/2020MNRAS.499.2303V} {499, 2303}

\bibitem[\protect\citeauthoryear{{Vall{\'e}s-P{\'e}rez}, {Planelles}  \&
  {Quilis}}{{Vall{\'e}s-P{\'e}rez} et~al.}{2021a}]{Valles_2021cpc}
{Vall{\'e}s-P{\'e}rez} D.,  {Planelles} S.,   {Quilis} V.,  2021a, \mn@doi
  [Computer Physics Communications] {10.1016/j.cpc.2021.107892}, \href
  {https://ui.adsabs.harvard.edu/abs/2021CoPhC.26307892V} {263, 107892}

\bibitem[\protect\citeauthoryear{{Vall{\'e}s-P{\'e}rez}, {Planelles}  \&
  {Quilis}}{{Vall{\'e}s-P{\'e}rez} et~al.}{2021b}]{Valles_2021}
{Vall{\'e}s-P{\'e}rez} D.,  {Planelles} S.,   {Quilis} V.,  2021b, \mn@doi
  [\mnras] {10.1093/mnras/stab880}, \href
  {https://ui.adsabs.harvard.edu/abs/2021MNRAS.504..510V} {504, 510}

\bibitem[\protect\citeauthoryear{{Vall{\'e}s-P{\'e}rez}, {Planelles}  \&
  {Quilis}}{{Vall{\'e}s-P{\'e}rez} et~al.}{2022}]{Valles_2022}
{Vall{\'e}s-P{\'e}rez} D.,  {Planelles} S.,   {Quilis} V.,  2022, \mn@doi
  [\aap] {10.1051/0004-6361/202243712}, \href
  {https://ui.adsabs.harvard.edu/abs/2022A&A...664A..42V} {664, A42}

\bibitem[\protect\citeauthoryear{{Vazza}, {Jones}, {Br{\"u}ggen}, {Brunetti},
  {Gheller}, {Porter}  \& {Ryu}}{{Vazza} et~al.}{2017}]{Vazza_2017}
{Vazza} F.,  {Jones} T.~W.,  {Br{\"u}ggen} M.,  {Brunetti} G.,  {Gheller} C.,
  {Porter} D.,   {Ryu} D.,  2017, \mn@doi [\mnras] {10.1093/mnras/stw2351},
  \href {https://ui.adsabs.harvard.edu/abs/2017MNRAS.464..210V} {464, 210}

\bibitem[\protect\citeauthoryear{{Villaescusa-Navarro}
  et~al.,}{{Villaescusa-Navarro} et~al.}{2021}]{Villaescusa-Navarro_2021}
{Villaescusa-Navarro} F.,  et~al., 2021, \mn@doi [\apj]
  {10.3847/1538-4357/abf7ba}, \href
  {https://ui.adsabs.harvard.edu/abs/2021ApJ...915...71V} {915, 71}

\bibitem[\protect\citeauthoryear{{Villaescusa-Navarro}
  et~al.,}{{Villaescusa-Navarro} et~al.}{2022}]{Villaescusa-Navarro_2022}
{Villaescusa-Navarro} F.,  et~al., 2022, arXiv e-prints, \href
  {https://ui.adsabs.harvard.edu/abs/2022arXiv220101300V} {p. arXiv:2201.01300}

\bibitem[\protect\citeauthoryear{{Virtanen} et~al.,}{{Virtanen}
  et~al.}{2020}]{Scipy}
{Virtanen} P.,  et~al., 2020, \mn@doi [Nature Methods]
  {10.1038/s41592-019-0686-2}, \href {https://rdcu.be/b08Wh} {17, 261}

\bibitem[\protect\citeauthoryear{{Voit}}{{Voit}}{2005}]{Voit_2005}
{Voit} G.~M.,  2005, \mn@doi [Reviews of Modern Physics]
  {10.1103/RevModPhys.77.207}, \href
  {https://ui.adsabs.harvard.edu/abs/2005RvMP...77..207V} {77, 207}

\bibitem[\protect\citeauthoryear{{Wang}, {Mao}, {Zentner}, {Lange}, {van den
  Bosch}  \& {Wechsler}}{{Wang} et~al.}{2020}]{Wang_2020}
{Wang} K.,  {Mao} Y.-Y.,  {Zentner} A.~R.,  {Lange} J.~U.,  {van den Bosch}
  F.~C.,   {Wechsler} R.~H.,  2020, \mn@doi [\mnras] {10.1093/mnras/staa2733},
  \href {https://ui.adsabs.harvard.edu/abs/2020MNRAS.498.4450W} {498, 4450}

\bibitem[\protect\citeauthoryear{{Weinberger}, {Springel}  \&
  {Pakmor}}{{Weinberger} et~al.}{2020}]{Weinberger_2020}
{Weinberger} R.,  {Springel} V.,   {Pakmor} R.,  2020, \mn@doi [\apjs]
  {10.3847/1538-4365/ab908c}, \href
  {https://ui.adsabs.harvard.edu/abs/2020ApJS..248...32W} {248, 32}

\bibitem[\protect\citeauthoryear{{Wetzel}, {Cohn}  \& {White}}{{Wetzel}
  et~al.}{2009}]{Wetzel_2009}
{Wetzel} A.~R.,  {Cohn} J.~D.,   {White} M.,  2009, \mn@doi [\mnras]
  {10.1111/j.1365-2966.2009.14424.x}, \href
  {https://ui.adsabs.harvard.edu/abs/2009MNRAS.395.1376W} {395, 1376}

\bibitem[\protect\citeauthoryear{{White} \& {Rees}}{{White} \&
  {Rees}}{1978}]{White_1978}
{White} S.~D.~M.,  {Rees} M.~J.,  1978, \mn@doi [\mnras]
  {10.1093/mnras/183.3.341}, \href
  {https://ui.adsabs.harvard.edu/abs/1978MNRAS.183..341W} {183, 341}

\bibitem[\protect\citeauthoryear{{Yu}, {Nelson}, {Nagai}  \& {Lau}}{{Yu}
  et~al.}{2014}]{Yu_2014}
{Yu} L.,  {Nelson} K.~L.,  {Nagai} D.,   {Lau} E.,  2014, in American
  Astronomical Society Meeting Abstracts \#223. p. 358.20

\bibitem[\protect\citeauthoryear{{Yuan}, {Han}  \& {Wen}}{{Yuan}
  et~al.}{2022}]{Yuan_2022}
{Yuan} Z.~S.,  {Han} J.~L.,   {Wen} Z.~L.,  2022, \mn@doi [\mnras]
  {10.1093/mnras/stac1037}, \href
  {https://ui.adsabs.harvard.edu/abs/2022MNRAS.513.3013Y} {513, 3013}

\bibitem[\protect\citeauthoryear{{Zel'dovich}}{{Zel'dovich}}{1970}]{Zeldovich_1970}
{Zel'dovich} Y.~B.,  1970, \aap, \href
  {https://ui.adsabs.harvard.edu/abs/1970A&A.....5...84Z} {5, 84}

\bibitem[\protect\citeauthoryear{{Zemp}, {Gnedin}, {Gnedin}  \&
  {Kravtsov}}{{Zemp} et~al.}{2011}]{Zemp_2011}
{Zemp} M.,  {Gnedin} O.~Y.,  {Gnedin} N.~Y.,   {Kravtsov} A.~V.,  2011, \mn@doi
  [\apjs] {10.1088/0067-0049/197/2/30}, \href
  {https://ui.adsabs.harvard.edu/abs/2011ApJS..197...30Z} {197, 30}

\bibitem[\protect\citeauthoryear{{Zhang}, {Cui}, {Dave}  \& {De
  Petris}}{{Zhang} et~al.}{2021a}]{ZhangB_2021}
{Zhang} B.,  {Cui} W.,  {Dave} R.,   {De Petris} M.,  2021a, arXiv e-prints,
  \href {https://ui.adsabs.harvard.edu/abs/2021arXiv211201909Z} {p.
  arXiv:2112.01909}

\bibitem[\protect\citeauthoryear{{Zhang}, {Zhuravleva}, {Kravtsov}  \&
  {Churazov}}{{Zhang} et~al.}{2021b}]{Zhang_2021}
{Zhang} C.,  {Zhuravleva} I.,  {Kravtsov} A.,   {Churazov} E.,  2021b, \mn@doi
  [\mnras] {10.1093/mnras/stab1546}, \href
  {https://ui.adsabs.harvard.edu/abs/2021MNRAS.506..839Z} {506, 839}

\makeatother
\end{thebibliography}




\appendix

\section{Evolution of the thresholds and weights on the dynamical state indicators for the high-mass sample}
\label{s:appA}

The high-mass subsample, defined in Sec. \ref{s:results.mass_dependence} and shown in Fig. \ref{fig6}, corresponds to a redshift-dependent mass limit, which can be parametrised by

\begin{equation}
    \log_{10} \frac{M_\mathrm{lim}(z)}{M_\odot} = 13.49 - 0.21 z
\end{equation}

\noindent within $\lesssim 0.05 \, \mathrm{dex}$. Thus, the sample corresponds to massive groups and clusters at $z\sim 0$; to objects above $10^{13} M_\odot$ at $z \sim 2$; and to a mass limit of $\sim 3 \times 10^{12} M_\odot$ at $z \sim 5$, which might most often be the progenitors of the massive haloes we find at $z \sim 0$.

Within this sample, the evolution with redshifts of the thresholds on the dynamical state indicators, shown in Fig. \ref{fig7}, can be given by the following polynomial fits:

\begin{equation}
    \Delta_r^\mathrm{thr}(z) \big|_\mathrm{massive} = 0.0863(39) + 0.0066(23) z 
    \label{eq:fit_centre_offset_massive}
\end{equation}
\begin{equation}
    \eta^\mathrm{thr}(z) \big|_\mathrm{massive}  = 1.3371(88) + 0.151(14) z - 0.0139(37) z^2 
    \label{eq:fit_virial_massive}
\end{equation}
\begin{equation}
    \langle \widetilde{v_r} \rangle_\mathrm{DM}^\mathrm{thr}(z) \big|_\mathrm{massive}  = 0.0842(32)
    \label{eq:fit_mean_vr_massive}
\end{equation}
\begin{equation}
    s_{200c,500c}^\mathrm{thr}(z) \big|_\mathrm{massive}  = 1.495(10)
    \label{eq:fit_sparsity_massive}
\end{equation}
\begin{equation}
    \varepsilon^\mathrm{thr}(z) \big|_\mathrm{massive}  = 0.2710(33)
    \label{eq:fit_ellipticity_massive}
\end{equation}

The weights on these indicators, as they appear on the relaxedness parameter (Eq. \ref{eq:xi_dynstate}), are fitted by:

\begin{equation}
    w[\Delta_r](z)\big|_\mathrm{massive} \propto 0.218(16) - 0.134(26) z + 0.0356(69) z^2 
    \label{eq:fit_centre_offset_weight_massive}
\end{equation}
\begin{equation}
    w[\eta](z) \big|_\mathrm{massive}\propto 0.250(11) - 0.0603(66) z 
    \label{eq:fit_virial_weight_massive}
\end{equation}
\begin{equation}
    w[\langle \widetilde{v_r} \rangle_\mathrm{DM}](z) \big|_\mathrm{massive} \propto 0.092(17) + 0.109(27) z - 0.0141(71) z^2 
    \label{eq:fit_mean_vr_weight_massive}
\end{equation}
\begin{equation}
    w[s_{200c,500c}](z) \big|_\mathrm{massive} \propto 0.2251(87)
    \label{eq:fit_sparsity_weight_massive}
\end{equation}
\begin{equation}
    w[\varepsilon](z) \big|_\mathrm{massive} \propto 0.2537(86)
    \label{eq:fit_ellipticity_weight_massive}
\end{equation}



\bsp	
\label{lastpage}
\end{document}